\documentclass{WileyMSP-template}

\usepackage{graphicx}
\usepackage{hyperref}
\hypersetup{plainpages=false,
pdfpagemode={UseOutlines},
bookmarksopen=false,
bookmarksopenlevel=0,
colorlinks=true,
citecolor=blue,% The color of citations
linkcolor=blue,% The color of references to document elements (sections, figures, etc)
urlcolor=blue,% The color of hyperlinks (URLs)
pdfstartview={FitV},
unicode,
breaklinks=true,
}

\usepackage{xcolor}
\usepackage{booktabs}
\usepackage{multirow}
\usepackage{amsmath}
\usepackage{amssymb}
\usepackage{array}
\usepackage{dcolumn}   % decimal-aligned table columns
\usepackage{textcomp}

\usepackage{parskip}
\usepackage{fancyhdr}
\fancypagestyle{nosection}{%
  \fancyhead[L]{}                       % drop the section mark
}

\usepackage{etoc}

\begin{document}

\pagestyle{fancy}
\rhead{\includegraphics[width=2.5cm]{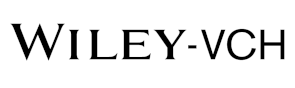}}

\title{Controlling the Sign and Magnitude of the Thermoreflectance Coefficient of Gold-MoS$_2$ Metamaterials}

\maketitle

% Author: Please give full first and last names for authors and include * after the name of all corresponding authors

\author{David Saleta Reig*}
\author{Bohai Liu}
\author{Kai Xu}
\author{Juan Sebastián Reparaz}
\author{Klaas-Jan Tielrooij*}

% Dedication

%\dedication{Optional dedication here. If no dedication is required, please leave blank}
\dedication{}

% Affiliations: Please provide adacemic titles (Prof. or Dr.) for all authors where applicable, and include an institutional email address for all corresponding authors
\begin{affiliations}
\mbox{Dr.~D.~Saleta Reig}, \mbox{Prof.~Dr.~K.-J.~Tielrooij}\\
Catalan Institute of Nanoscience and Nanotechnology (ICN2), CSIC and BIST, Campus UAB, 08193 Bellaterra, Barcelona, Spain\\
\medskip
\mbox{Dr.~K.~Xu}, \mbox{Prof.~Dr.~J.~S.~Reparaz}\\
Institut de Ciència de Materials de Barcelona, ICMAB-CSIC, Campus UAB, 08193 Bellaterra, Spain\\
\medskip
\mbox{Dr.~D.~Saleta~Reig}, \mbox{Dr.~B.~Liu}, \mbox{Prof.~Dr.~K.-J.~Tielrooij}\\
Department of Applied Physics, TU Eindhoven, Den Dolech 2, Eindhoven, 5612 AZ, Netherlands\\
\medskip
Email Addresses: \href{mailto:david.saleta@icn2.cat}{david.saleta@icn2.cat}; \href{mailto:k.j.tielrooij@tue.nl}{k.j.tielrooij@tue.nl}\\

\end{affiliations}

% Keywords: Please provide a minimum of three and a maximum of seven keywords, separated by commas


% Abstract should be written in the present tense and impersonal style (i.e., avoid we), and be at most 200 words long
\begin{abstract}

Heating a material changes how much light it reflects. This effect, quantified by the thermoreflectance coefficient $C_\mathrm{TR}$, enables temperature sensing, while it is a nuisance for situations where constant reflection is crucial. Whereas temperature detection requires a large $C_\mathrm{TR}$, the latter situation demands it to be as close to zero as possible. Controlling the $C_\mathrm{TR}$ has so far relied on the selection of materials and wavelengths, rather than engineering the coefficient itself. This work demonstrates that the $C_\mathrm{TR}$ of a 50~nm thin gold film on SiO$_2$/Si can be controlled in sign and magnitude by an ultrathin ($<$8~nm) MoS$_2$ film underneath it. Time-resolved thermoreflectance measurements reveal suppression, sign inversion, and revival of the thermoreflectance signal, controlled by three knobs: (\textit{i})~the buried MoS$_2$ thickness, (\textit{ii})~probe wavelength, and (\textit{iii})~ambient temperature. Transfer-matrix simulations using temperature-dependent dielectric functions of gold and MoS$_2$ reproduce all experimental observations, demonstrating an optothermal metamaterial platform with controllable properties. In particular, these (anti-)thermoreflective metamaterial mirrors bring $C_\mathrm{TR}$ under the same interference design rules previously applied to the optical response itself, opening ways to maximize or minimize the temperature dependence of the optical reflectance, tailored to the specific application.

\end{abstract}

% Text: Please use section headings and subheadings as specified below. For communications, all section headings apart from Experimental Section should be removed
% Please make the first reference to a display item bold: \textbf{Figure 1}
% Do not abbreviate Figure, Equation, etc.; display items are always singular, i.e., Figure 1 and 2.
% Equations are always singular, i.e., Equation 1 and 2, and should be inserted using the {equation} environment, not as graphics
% Please do not use footnotes in the text, additional information can be added to the Reference list.
\etocdepthtag.toc{main}
% Introduction section ----------------------------------------------------
\section{Introduction}

The optical reflectance $R$ of a surface generally depends on its temperature $T$ as quantified by the thermoreflectance coefficient \mbox{($C_\mathrm{TR} \equiv R^{-1}\partial R/\partial T$)}, which is the reflectivity change per unit Kelvin. This dependence is increasingly exploited as a control knob to reconfigure optical responses thermally~\cite{thermo_optics_NanoLetters2026}. In some cases, the $C_\mathrm{TR}$ should ideally be as close to zero as possible. For example, in phase-change photonic memories and neuromorphic hardware, the reflectance that encodes a non-volatile material state drifts with transient self-heating during read and write operations, so that a structural change and an ordinary temperature rise are reported through the same number~\cite{Wuttig2017,Nobile2023}. In precision interferometry, temperature fluctuations in a mirror coating couple into the measurement through thermoelastic and thermorefractive effects and set a noise floor, for example in gravitational-wave detectors and optical reference cavities~\cite{Evans2008}.

A large $C_\mathrm{TR}$ is desirable in optical thermometry and heat metrology, where thin metal films often serve as transducers that convert a temperature excursion into a measurable change in reflected light~\cite{Cahill2004,Schmidt_FDTR2009,Cahill2014,jiang2018tutorial}. Even for the metals of choice, the coefficient is modest (\mbox{$|C_\mathrm{TR}| \sim 10^{-5}$--$10^{-3}$~K$^{-1}$}), and peaks only in narrow spectral windows. Therefore, in practice, each metal transducer is used in combination with a specific probe wavelength~\cite{Wilson2012_thermoreflecCoeff}. Gold is a popular transducer material, which has a $C_\mathrm{TR}$ spectrum that crosses zero at the interband transition ($\sim$2.5~eV) and peaks with opposite signs on either side, reaching \mbox{$C_\mathrm{TR} \approx -3\times10^{-4}$~K$^{-1}$} at 515~nm, the largest value among metals~\cite{Wilson2012_thermoreflecCoeff,CTR_vs_T_metals_2015}. 

However, whether the goal is sensitivity or stability, $C_\mathrm{TR}$ is generally treated as a fixed material property for a given wavelength: its magnitude and spectral dependence are set by the material's optical response, leaving little room for control and design. 
Two lines of work have achieved some degree of control over $C_\mathrm{TR}$. First, in thermoreflectance metrology, layers added on top of the metal transducer are known to change the coefficient through interference within the added film, first as a calibration complication introduced by device passivation~\cite{Farzaneh2009}. Later, a deliberate choice of dielectric thickness was used to improve the signal-to-noise ratio~\cite{CTR_vs_T_metals_2015}; as well as by wetting the surface with a liquid, which enhances the coefficient by more than an order of magnitude~\cite{Alajlouni2022}. Second, in precision optics, the layer structure of crystalline Bragg mirrors has been optimized so that thermoelastic and thermorefractive contributions cancel coherently, suppressing thermo-optic noise~\cite{Gorodetsky2008,Chalermsongsak2016}. The first approach acts from above the metal, at thicknesses comparable to the probe wavelength, and targets sensitivity alone. The second nulls a fluctuation in reflected phase, distributed across many quarter-wave periods of a dielectric stack. However, neither of these two approaches offers deliberate control over the sign and amplitude response $\partial R/\partial T$ of a metal surface.

Here, we show that we can achieve this using a single buried layer with deep-subwavelength thickness, based on a two-dimensional (2D) semiconductor of the transition-metal dichalcogenide (TMD) family, namely MoS$_2$. The sharp excitonic resonances of TMDs~\cite{beal1979kramers,Ermolaev2020} shift and broaden strongly with temperature, at rates on the order of $10^{-4}$~eV~K$^{-1}$~\cite{Liu2020}. As a result, few-nanometer TMD films reach thermo-optic coefficients that exceed those of many metals by up to an order of magnitude in the visible~\cite{Varghese2026}. At the same time, their high refractive index \mbox{($n \sim 4$--$5$)}~\cite{Ermolaev2021,Munkhbat2022}, strong absorption, and self-passivated, dangling-bond-free surfaces allow van der Waals integration onto virtually any surface with sub-nanometer thickness control. These properties are already known to make TMDs exceptional optical interference layers: films thinner than 15~nm placed on a metal back-reflector sustain strong interference despite their deep-subwavelength thickness, allowing for the creation of optical metamaterials reaching near-unity absorption~\cite{Kats2013,Kats2016,unityAbs_TMDs_NanoLett2016,Munkhbat2019}. We demonstrate that the same properties make them equally exceptional interference layers for thermo-optics by burying the TMD beneath the metal to engineer the temperature derivative of the metal's reflection coefficient $\partial R/\partial T$, thus creating optothermal metamaterials.

Specifically, using Au(50~nm)/Ti(2~nm)/MoS$_2$(0--8~nm) on SiO$_2$(285~nm)/Si substrates, our time-resolved optothermal measurements show that nanometer-scale changes in MoS$_2$ thickness suppress, reverse the sign of, and revive the thermoreflectance signal, consistently across all pump-probe delays. Here, the sub-wavelength stack responds as a single effective surface with a thermoreflectance coefficient of its own: an ``optothermal metamaterial mirror''.  
Besides thickness, probe wavelength and ambient temperature provide two additional control knobs, and all three dependencies are reproduced by transfer-matrix method (TMM) simulations employing temperature-dependent dielectric functions of Au and MoS$_2$. Programming the sign and magnitude of a metal's $C_\mathrm{TR}$ using a single buried nanometer-scale layer extends interference coating design from the optical response to the optothermal response.

% Results section ---------------------------------------------------------

\section{Results and Discussion}

\begin{figure}[htbp]
  \centering
  \includegraphics[width=0.85\linewidth]{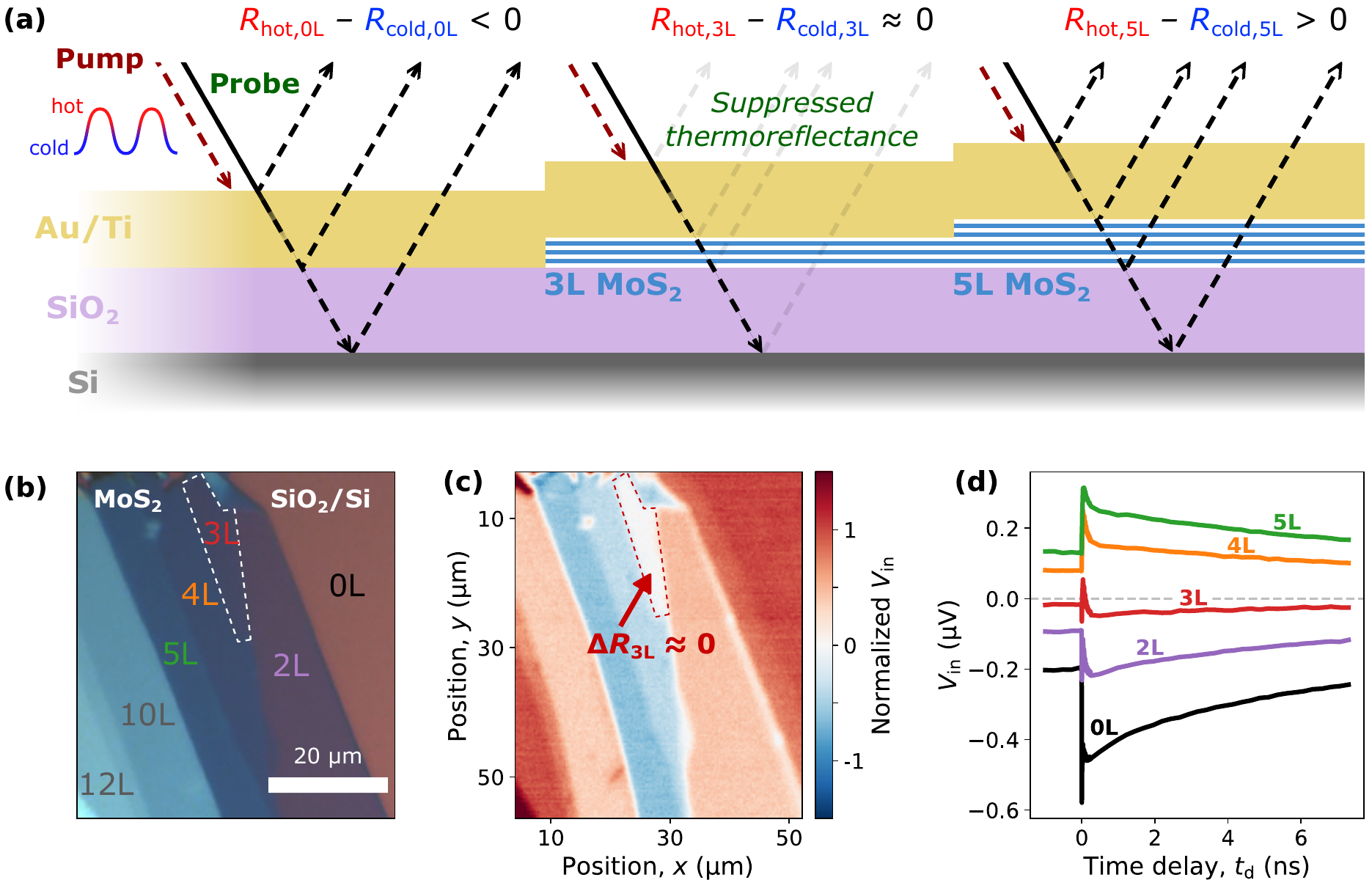}
  \caption{\textbf{Thickness-dependent thermoreflectance of Au-MoS$_2$ metamaterial mirrors at room temperature.} \textbf{(a)}~Side view schematic of the pump-probe optothermal experiments on a Au/Ti/MoS$_2$/SiO$_2$/Si sample with varying MoS$_2$ thickness. The pump beam (\textit{dashed red arrow}), whose amplitude is modulated at MHz frequencies, induces dynamic temperature changes in the metamaterial mirror. The probe beam (\textit{solid black line}) detects the pump-induced reflectivity changes at the modulation frequency of the pump beam. The multiple internal reflections of the probe beam (\textit{dashed black arrows}) suppress the effective thermoreflectance of the metamaterial mirror with 3L MoS$_2$ \mbox{($\Delta R_\mathrm{3L} \approx 0$)}. \textbf{(b)}~Top view optical image of the terrace-like MoS$_2$ flake prior to deposition of the \mbox{$50+2$~nm} Au/Ti film. \textbf{(c)}~Thermoreflectance map normalized to the signal magnitude of the 0L region, obtained using spatially overlapped pump and probe beams at a time delay ($t_d$) of 100~ps. The thermoreflectance signal turns again positive for MoS$_2$ flakes thicker than 10L. \textbf{(d)}~In-phase lock-in signal ($V_\mathrm{in}$) for various MoS$_2$ flake thickness, showing a cross-over from negative sign when MoS$_2$ is absent (0L) to positive sign ($>$3L MoS$_2$). Data of panels (c) and (d) obtained using 515~nm probe wavelength ($\lambda_\mathrm{pr}$), 1.11~MHz modulation frequency ($f_\mathrm{mod}$) of the pump beam, and $2$~\textmu m average spotsizes ($w_0$).}
  \label{fig:RT_thermoreflectance_Figure1}
\end{figure}

To fabricate Au/Ti/MoS$_2$ stacks on SiO$_2$/Si substrates, we use dry-transfer of MoS$_2$ flakes and e-beam evaporation of Au/Ti, as described in Section~\ref{sec:SampleFab}. To exploit the multilayer nature of our samples and enable thermoreflectance control (see \textbf{Figure~\ref{fig:RT_thermoreflectance_Figure1}a}), we choose a gold film thickness of 50~nm, which is comparable to its optical penetration depth (20--45~nm) in the visible~\cite{Olmon2012}. To study how the $C_\mathrm{TR}$ depends on the buried MoS$_2$ thickness, we use mechanical exfoliation to produce terrace-like MoS$_2$ flakes: flakes exhibiting discrete stepwise thickness variations, where adjacent regions differ by a single MoS$_2$ layer thickness ($0.615$~nm). With this approach, we identify multiple regions ranging from 12 layers (12L) to the monolayer (1L) limit within three terrace-like MoS$_2$ samples (see Figure~\ref{fig:RT_thermoreflectance_Figure1}\textbf{b}, \textbf{\ref{fig:SI_OpticalContrast}a--c} and \textbf{\ref{fig:FIG_S-samples}a--c}). 

We perform pump-probe thermoreflectance imaging (at a fixed time delay) and time-resolved optothermal measurements on our MoS$_2$-thickness-controlled metamaterial mirrors, using a time-domain thermoreflectance setup (see \textbf{Figure~\ref{fig:TDTRsetup}}). 
A pump beam modulated at \mbox{$f_\mathrm{mod} = 1.11$~MHz} (unless otherwise specified) induces small dynamic temperature changes, and the effective $C_\mathrm{TR}$ determines the sign and magnitude of the resulting reflectance change (\mbox{$\Delta R/R = C_\mathrm{TR}\Delta T$}). A spatially overlapped probe pulse detects this change at a controllable time delay $t_\mathrm{d}$ between pump \mbox{($\lambda_\mathrm{pu} = 1030$~nm)} and probe \mbox{($\lambda_\mathrm{pr} = 515$~nm)} pulses. We collect the reflected probe light on a silicon photodiode connected to a lock-in amplifier, which demodulates the signal at $f_\mathrm{mod}$ to isolate the temperature-induced changes in reflectance. 

We exploit the lock-in signal to infer the thermoreflectance coefficient of our transducer (meta)material. In what follows, we examine the thermoreflectance signals as a function of the three parameters that set the operating point of the metamaterial mirror: ($i$)~the MoS$_2$ thickness, which fixes the interference condition at fabrication (Section~\ref{sec:thickness_tunable}); ($ii$)~the probe wavelength, which selects where on the resulting thermoreflectance spectra the readout occurs (Section~\ref{sec:Probe_wavelength}); and ($iii$)~the ambient temperature, which shifts the underlying dielectric functions and thereby tunes the response without any change to the structure (Section~\ref{sec:Temperature_control}). Together, the three dependencies provide complementary routes for thermoreflectance control by exploiting interference from the multiple internal reflections within the sample, and all are captured by the same transfer-matrix model. 

\subsection{Thickness-tunable and sign-reversible effective thermoreflectance}
\label{sec:thickness_tunable}

Figure~\ref{fig:RT_thermoreflectance_Figure1}\textbf{c} presents an exemplary thermoreflectance map of the in-phase lock-in signal, retrieved from the same sample region shown in Figure~\ref{fig:RT_thermoreflectance_Figure1}\textbf{b} after the 50~nm gold layer has been deposited. The thermoreflectance images reveal an unexpectedly large contrast between the sample regions with buried MoS$_2$ of different thicknesses --with, surprisingly, single-layer resolution. Here, some regions barely show any signal (3L), and some regions show sign-reversed thermoreflectance signal (4L and 5L) with respect to that of the region without MoS$_2$ (0L). We proceed with acquiring thermoreflectance maps from additional samples (see examples in Figure~\ref{fig:FIG_S-samples}\textbf{d--f}) and confirm the consistent suppression of thermoreflectance from a specific MoS$_2$ thickness (3L at room temperature), while noticing a second sign-reversal for thicker MoS$_2$ layers ($\sim$9L at room temperature). In addition, we observe enhanced visibility to cracks, wrinkles and imperfections in the samples (see \textbf{Figure~\ref{fig:FIG_S-cracks}}). 

We use such thermoreflectance maps to choose the regions of interest for further time-resolved optothermal measurements. In Figure~\ref{fig:RT_thermoreflectance_Figure1}\textbf{d}, we present exemplary time-resolved data, which clearly reproduce our observations from thermoreflectance imaging. Introducing few-layer MoS$_2$ flakes at the metal-substrate interface produces a progressive sign reversal in discrete steps when increasing the number of buried MoS$_2$ layers. At room temperature, the magnitude of the time-resolved thermoreflectance signal decreases when adding MoS$_2$ layers, eventually becoming minimal for 3L MoS$_2$, and changing sign for 4L and 5L with respect to the region without MoS$_2$ (0L). This implies that, almost independently of the temporal delay between pump and probe pulses, we can achieve single-layer resolution, differentiating the distinct MoS$_2$ regions in our sample using both thermoreflectance imaging and time-resolved optothermal measurements.

In Section~\ref{sec:robust_TR} we show that our results are independent of analyzing either the in-phase or the out-of-phase lock-in signals (see \textbf{Figure~\ref{fig:VinVout_sametrend}}), the laser spot sizes used (see \textbf{Figure~\ref{fig:SI-FIG_RTpos_SignalMagnitude}}), and the modulation frequency of the pump beam (see \textbf{Figure~\ref{fig:SI-FIG_fmod_SignalMagnitude}}). Furthermore, we support the results using frequency domain thermoreflectance (FDTR) measurements (see Section~\ref{sec:FDTR}). We therefore attribute our findings to a modified effective thermoreflectance coefficient $C_\mathrm{TR}$ mediated by the presence of interfacial MoS$_2$, whose sub-nm thickness control allows for tailoring interference from the multiple internal reflections of the optothermal field at the probe wavelength.

\subsection{Probe-wavelength control of the effective thermoreflectance}
\label{sec:Probe_wavelength}

\begin{figure}[ht]
    \centering
  \includegraphics[width=\linewidth]{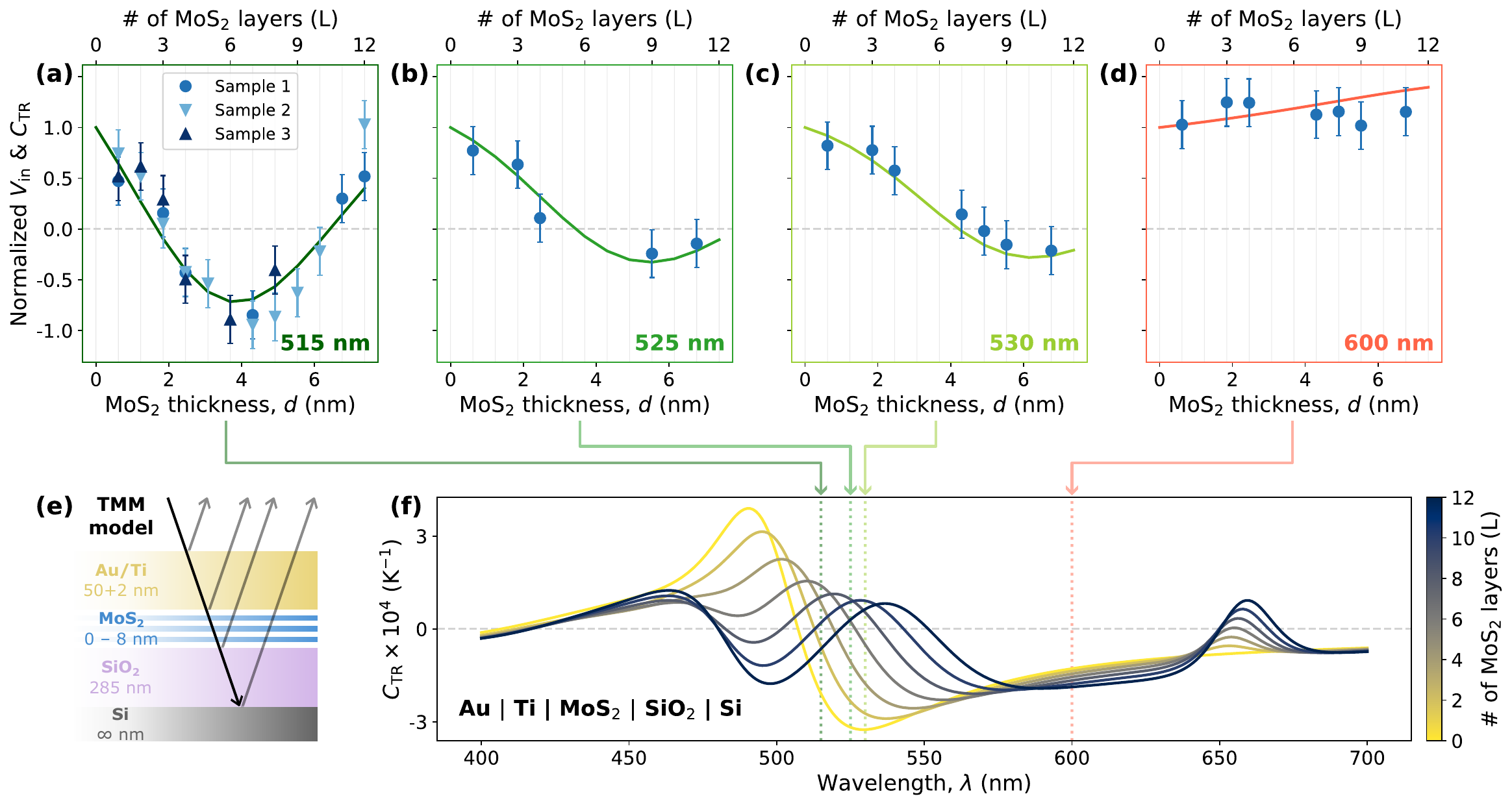}
  \caption{\textbf{Wavelength-dependent optothermal experiments and simulations.} Experimental in-phase ($V_\mathrm{in}$) signal magnitudes at nanosecond time-delays and normalized to the signal magnitude of the 0L region (\textit{scatters}); using \textbf{(a)}~515~nm, \textbf{(b)}~525~nm, \textbf{(c)}~530~nm, and \textbf{(d)}~600~nm probe wavelength; and corresponding transfer-matrix method (TMM) simulations (\textit{solid lines}). The error bars represent the largest experimental uncertainty of $\sim$24$\%$ obtained across repeated measurements within a given sample region, which we then applied to all experimental data points. \textbf{(e)}~Side view scheme of the sample geometry used in the TMM simulations, depicting the multiple internal reflections of the incident probe light. \textbf{(f)}~Thermoreflectance coefficient retrieved from our wavelength-resolved TMM simulations of a Au/Ti/MoS$_2$/SiO$_2$/Si sample with MoS$_2$ thickness varied from 0 to 12L.}
  \label{fig:CTRvsWvl_Figure2}
\end{figure}

To confirm that our results can be attributed to optical interference phenomena, we proceed with studying the effects of probe wavelength on the thermoreflectance measured from our Au-MoS$_2$ metamaterial mirrors. We thus perform measurements at four different probe wavelengths --515, 525, 530 and 600~nm-- and analyze the thermoreflectance signal magnitude at time delays of \mbox{$t_\mathrm{d}>3$~ns}, to ensure that the sample has thermalized. We present these results in \textbf{Figure~\ref{fig:CTRvsWvl_Figure2}a--d}, where we observe significant differences in the MoS$_2$-thickness-dependent trends when varying the probe wavelength. At \mbox{$\lambda_\mathrm{pr} = 515$~nm}, the thermoreflectance signal magnitude decreases with the introduction of MoS$_2$ layers, reaching a minimum value for a region with 3L MoS$_2$. Further increasing the number of layers reverses the sign of the thermoreflectance signal reaching a similar base level of sign-reversed signal for 6L MoS$_2$. The signal magnitude then decreases again and becomes minimum for $\sim$9L MoS$_2$, while recovering the original sign for thicker flakes. At slightly longer probe wavelengths of 525 and 530~nm, we observe the first zero-crossing points to occur for thicker MoS$_2$ regions: for 4 and 8 layers, respectively. Interestingly, all measurements on Au-MoS$_2$ regions show a magnitude similar to the control region without MoS$_2$ when using \mbox{$\lambda_\mathrm{pr} = 600$~nm}. This dependence on probe wavelength is consistent with the hypothesis that optical interferences play an important role. 

In order to understand how optical interferences modify the thermal reflectance coefficient, we perform transfer-matrix method (TMM) simulations (see Section~\ref{subsec:TMM_methods} and \ref{sec:tmm}) that take into account the wavelength-dependent optical properties of gold and MoS$_2$. We retrieved these properties from a Drude-Lorentz model for gold and from a two-Lorentz excitonic oscillator model for MoS$_2$. For simplicity, we considered the optical properties of Ti, SiO$_2$ and Si to be constant within our experimental wavelength range (see Section~\ref{subsec:TMM_methods}). Before presenting our results, we note that our TMM simulations of the whole stack without MoS$_2$ agree with simulated~\cite{Au_model_FarbaniecEakins2023} and experimental~\cite{CTR_howMeasureTeTph_2022} $C_\mathrm{TR}$ values reported in the literature. 

In Figure~\ref{fig:CTRvsWvl_Figure2}\textbf{f} we show the TMM-simulated thermoreflectance spectra for varying MoS$_2$ thicknesses. Clearly, introducing MoS$_2$ at the metal-substrate interface has a strong effect on the total thermoreflectance spectra only when the gold is sufficiently thin (see \textbf{Figure~\ref{fig:AuThickness_Sim}a} and~\ref{fig:AuThickness_Sim}\textbf{b}). Here, the presence of MoS$_2$ appears to have an opposing contribution to the $C_\mathrm{TR}$ spectra with respect to that of gold. The originally positive lobe of the $C_\mathrm{TR}$ of gold centered at $\sim$490~nm diminishes and becomes negative when adding MoS$_2$ layers; and the negative lobe centered at $\sim$525~nm starts to become positive. Importantly, the zero-crossing point between the two lobes further red-shifts when increasing the MoS$_2$ thickness, bringing the zero-crossing point past our probe wavelength. This is the main reason for the observed sign-reversal, and for the controlled suppression of thermoreflectance with a specific number of MoS$_2$ layers. 

Our simple TMM simulations describe our experimental data at all probe wavelengths used, suggesting that the MoS$_2$ flakes underneath 50~nm gold contribute more strongly to the total thermoreflectance signal at shorter green wavelengths. Also, our simulations reproduce the experimentally observed trends with MoS$_2$ thickness. Within the explored range, we find two zero-crossing points at 515~nm (for 3L and 9L), and one at 525 (5--6L) and 530~nm (6--7L), all of which coincide with nanometer thickness precision ($<$2 MoS$_2$ layers) with our experimental findings. At 600~nm, we find a constant $C_\mathrm{TR}$ when varying the buried flake thickness, in agreement with our experiments. Moreover, our simulations capture the signal magnitudes at all wavelengths within experimental uncertainty, including the sign-reversed and sign-recovered magnitudes. 

These observations are only possible because the thickness of our gold films is comparable to the optical penetration depth, which we calculate to be around $\sim$40~nm at \mbox{$\lambda = 515$~nm} using our TMM model (see \textbf{Figure~\ref{fig:penetrationDepth}}). Our results demonstrate that the thermoreflectance coefficient $C_\mathrm{TR}$ of thin gold films can be effectively engineered using interfacial MoS$_2$. For example, suppressing thermoreflectance with flakes thinner than 5~nm at green wavelengths, thus enabling the design of metamaterial mirrors with small (large) thermoreflectance coefficients at the desired probe wavelength in the visible range.

\subsection{Temperature control and co-tuning with thickness: suppression and sign reversal}
\label{sec:Temperature_control}

\begin{figure}[ht]
    \centering  \includegraphics[width=1\linewidth]{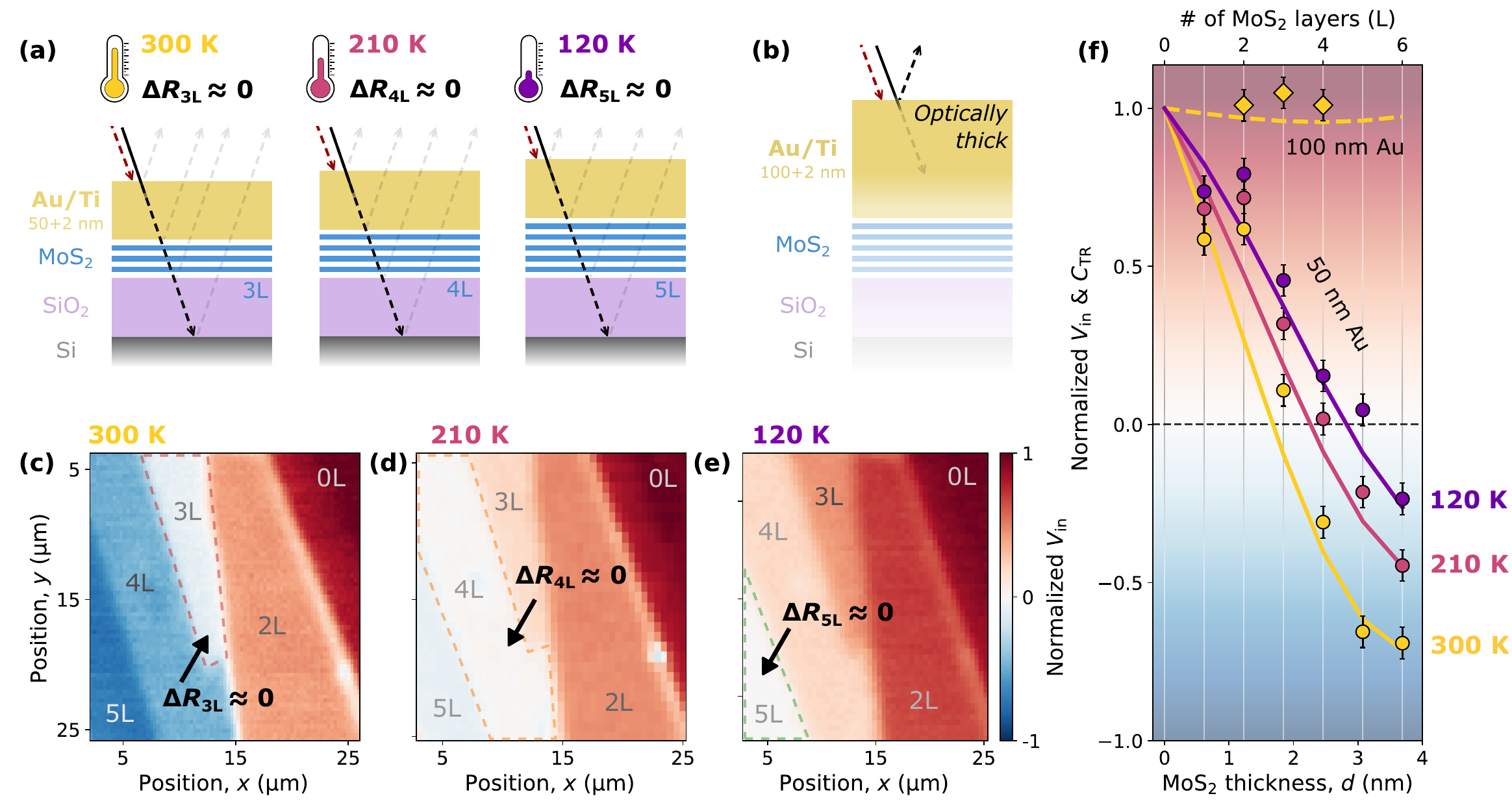}
  \caption{\textbf{Temperature-dependent optothermal experiments and simulations.} \textbf{(a)}~Side view scheme of the controllable suppression of thermoreflectance in 50~nm Au-MoS$_2$ metamaterial mirrors with ambient temperature. \textbf{(b)}~Side view scheme of a sample using 100~nm Au, where the gold is optically thick and the thermoreflectance is dominated by gold.
  \textbf{(c--e)}~Exemplary thermoreflectance maps normalized to signal magnitude of the 0L region, obtained at \textbf{(c)}~300~K, \textbf{(d)}~210~K, and \textbf{(e)}~120~K, where the thermoreflectance signal is minimum for 3L, 4L and 5L MoS$_2$, respectively. \textbf{(f)}~Summary of the experimental (\textit{scatters}) and simulated (\textit{lines}) signal magnitude scaling at 300~K (\textit{yellow}), 210~K (\textit{magenta}), and 120~K (\textit{purple}), using 50~nm Au (\textit{circles}) and 100~nm Au (\textit{diamonds}). The experimental data points correspond to the mean signal magnitude at long time delays \mbox{($t_\mathrm{d}>10$ ns)}, obtained from time-resolved experiments averaged from two terrace-like MoS$_2$ flakes coated with 50~nm gold, and from a single terrace-like flake coated with 100~nm gold. The largest experimental uncertainty across these measurements is $\sim$4$\%$.}
  \label{fig:TemperatureDep_Figure3}
\end{figure}

We will now proceed to investigate the effects of ambient temperature on the optothermal properties of our Au-MoS$_2$ metamaterials, offering an additional degree of freedom for tuning the $C_\mathrm{TR}$ of our samples. In \textbf{Figure~\ref{fig:TemperatureDep_Figure3}} we show our temperature-dependent optothermal measurements. Here, we repeat measurements in regions with few-layer MoS$_2$ flakes ($<$6L) using \mbox{$\lambda_\mathrm{pr} = 515$~nm} from 120~K to room temperature in 30~K increments. Our temperature-dependent thermoreflectance imaging reveals that $C_\mathrm{TR}$ is significantly suppressed for MoS$_2$ thicknesses of 3L at 300~K, 4L at 210~K, and 5L at 120~K (see Figure~\ref{fig:TemperatureDep_Figure3}\textbf{a} and Figure~\ref{fig:TemperatureDep_Figure3}\textbf{c--e}), with the ambient temperature providing a fine tuning knob for suppressed thermoreflectance conditions of 1~layer (0.615~nm) thickness change per $\sim$90~K. We confirm this picture using time-resolved optothermal measurements at each temperature (see \textbf{Figure~\ref{fig:MinimumSignals_vs_T}} and~\ref{fig:TemperatureDep_Figure3}f).

Our TMM simulations take into account the temperature via linear thermo-optic coefficients, which we applied to the parameters of both Au and MoS$_2$ (see \textbf{Table~\ref{tab:params}}). We have seen from our simulations that the thermoreflectance of thicker gold films of 100~nm do not show any significant contribution from MoS$_2$ flakes placed underneath, and the thermoreflectance spectra remain unaffected (see Figure~\ref{fig:AuThickness_Sim}\textbf{b}). We confirm this result experimentally, by measuring a sample with 100~nm gold deposited on a terrace-like MoS$_2$ flake (see a side-view schematic of the sample in Figure~\ref{fig:TemperatureDep_Figure3}\textbf{b}). As expected, we do not see any suppression of thermoreflectance when introducing atomically thin MoS$_2$, as opposed to the effectively suppressed thermoreflectance using 3L MoS$_2$ at room temperature observed using 50~nm gold films. For thin gold, our simulations also capture the experimental MoS$_2$ thicknesses at which zero-crossings and sign reversal occur, with less than one layer accuracy for all temperatures (see Figure~\ref{fig:TemperatureDep_Figure3}\textbf{f}). 

% Conclusions section -----------------------------------------------------

\section{Conclusion}

We have systematically studied the optothermal properties of Au-MoS$_2$ metamaterial mirrors. We explored the effects of MoS$_2$ thickness, with sub-nm thickness control, and the influence of probe wavelength and ambient temperature on our experiments and simulations. 
Both our simulations (see Figure~\ref{fig:AuThickness_Sim}) and experiments (see Figure~\ref{fig:TemperatureDep_Figure3}\textbf{f}) reveal that thermoreflectance control is possible when gold is optically thin and the material underneath (MoS$_2$) also contributes to the total thermoreflectance. This enables the use of multiple internal reflections in the sample to further engineer thermoreflectance properties for a wide range of applications (\textit{e.g.}~in sensing). First, we have identified a sign-reversible thermoreflectance coefficient when using 50~nm gold films upon controlling the number of MoS$_2$ layers underneath, with specific MoS$_2$ thicknesses canceling the effective thermoreflectance from the sample. At room temperature, we observed a strong influence of the probe wavelength, which controls the thickness at which we observe the sign reversal. Second, the thickness at which sign reversal occurs shifts from 3L at 515~nm to 7L at 530~nm; while at 600~nm, no sign reversal occurs for MoS$_2$ thicknesses between 1 and 12L. Third, at 515~nm probe wavelength, decreasing the absolute temperature progressively shifted the MoS$_2$ thickness at which sign-reversal occurs, from 3L at 300~K to 5L at 120~K. Our TMM simulations also reproduced the temperature-dependent optothermal experiments when considering linear thermo-optic coefficients for Au and MoS$_2$. Such a degree of control makes Au-MoS$_2$ metamaterials a versatile platform for designing mirrors with an engineered thermoreflectance coefficient, suitable for multiple applications ranging from phonon sensing to anti-thermo-reflective metamaterial surfaces for optothermal camouflage.

% Experimental section ----------------------------------------------------

\section{Experimental Section}

\subsection{Sample fabrication}
\label{sec:SampleFab}
Mechanical exfoliation of bulk MoS$_2$ (2H phase, HQ Graphene) was used to produce thin flakes on polydimethylsiloxane (PDMS). Three terrace-like MoS$_2$ flakes (\textit{i.e.},~flakes consisting of adjacent regions with discrete step-like thickness variations) were identified under an optical microscope to cover the entire range of thickness from 1L to 12L MoS$_2$. The thickness of the MoS$_2$ regions of interest was determined using optical contrast and atomic force microscopy for thin and thick flakes, respectively (see Section~\ref{sec:SI_SampleFabCharact}). The terrace-like MoS$_2$ flakes were then transferred onto 285~nm SiO$_2$/Si substrates (University Wafer) using a dry-transfer approach described in Ref.~\cite{varghese2021fabrication}. Prior to the pump-probe optothermal experiments, a 50~nm gold film with 2~nm titanium adhesion layer was deposited on all samples using e-beam evaporation (FC-2000, Temescal). A fourth terrace-like MoS$_2$ flake was transferred onto SiO$_2$/Si and coated with 100~nm gold to study the influence of gold thickness.

\subsection{Pump-probe optothermal experiments}
\label{sec:OptothermalMethods}
A fundamental laser source (FLINT, Light Conversion) emitting $\sim$100~fs laser pulses centered at 1030~nm, at a repetition rate of 76~MHz, was used to generate the pump beam. A beamsplitter produced a second beam, which was frequency doubled using a non-linear lithium triborate crystal (LBO-603H, Eksma Optics) to obtain a 515~nm probe beam. The pump beam passed through an electro-optic modulator (EO-AM-NR-C2, Thorlabs) producing a sinusoidal amplitude modulation in the \mbox{$0.1$--$10$~MHz} range, and over a retroreflector-based 325~mm quadruple-pass mechanical delay line (DL-BKIT4U-S-M, Newport) to introduce a controlled time delay between pump and probe pulses. Both beams recombined on a harmonic beamsplitter (HBSY12, Thorlabs) before focusing on the sample using an objective lens (50X Plan Apo NIR Infinity Corrected, Mitutoyo). The transducer material linearly translates small lattice temperature changes \mbox{($\Delta T_\mathrm{l} = Q/C_\mathrm{l}<10$~K)}, given by the heat flux $Q$ and the lattice heat capacity $C_\mathrm{l}$, into measurable differences in reflectance~\cite{Cahill2004,jiang2018tutorial}:
\begin{equation}
    \Delta R/R = \left( \frac{1}{R} \frac{\partial R}{\partial T} \right) \cdot \Delta T = C_\mathrm{TR} \cdot \Delta T
\end{equation}
with $C_\mathrm{TR}$ its thermoreflectance coefficient. The reflected probe light was collected on a cathode-grounded silicon photodiode (SM05PD2A, Thorlabs) connected to a lock-in amplifier (HF2LI, Zurich Instruments), used to demodulate the signal at the modulation frequency of the pump beam. All time-resolved in-phase ($V_\mathrm{in}$) and out-of-phase ($V_\mathrm{out}$) lock-in signals shown in this manuscript were first corrected to account for the instrumentation phase offset, by rotation in the complex plane with the standard procedure used in time-domain thermoreflectance data analysis~\cite{Cahill2004, jiang2018tutorial}. All measurements were performed at best-focusing conditions \mbox{($w_0 = 2.2$~\textmu m)}, using \mbox{$f_\mathrm{mod} = 1.11$~MHz} and \mbox{$\lambda_\mathrm{probe} = 515$~nm}, and at room temperature (300~K), unless otherwise specified (more details in Section~\ref{sec:Setup}). 

    \subsubsection{Analysis of the thermoreflectance signal magnitude}
    \label{sec:AnalysisMethods}
    For $V_\mathrm{in}$, we first masked the early time delays of the signal (from \mbox{$t_\mathrm{d} =-50$~ps} to \mbox{$t_\mathrm{d} = 3$~ns}). We took the long time delays for each measurement (on a region with MoS$_2$) and normalized it with respect to the control signal (on a region without MoS$_2$). As the normalized signals were nearly constant with time delay, we averaged them to obtain a normalized signal magnitude that could be compared across different measurements and samples, each using its own control. This differential approach allowed us to remove the contributions from variable experimental conditions and to purely asses the influence of introducing MoS$_2$ flakes at the transducer-substrate interface. 

    \subsubsection{Wavelength-dependent optothermal experiments}
    The wavelength-dependent measurements shown in Figure~\ref{fig:CTRvsWvl_Figure2} involved using a different probe beam, generated with the combination of various non-linear stages. The fundamental beam passed through an optical parametric oscillator, whose signal was fed to a third-harmonic conversion stage to obtain 515, 525, 530 or 600~nm probe wavelength (see Figure~\ref{fig:TDTRsetup}). The optical path length of the new probe beam was adjusted to match the optical path length of our pump beam.
    
    \subsubsection{Temperature-dependent optothermal experiments}
    For the temperature-dependent measurements shown in Figure~\ref{fig:TemperatureDep_Figure3}, the temperature was varied from 120~K to 300~K in 30~K increments. At each temperature, the sample was allowed to thermalize for a minimum of 30~min prior to data acquisition.

\subsection{Transfer-matrix method (TMM) simulations}
\label{subsec:TMM_methods}

Effective thermoreflectance spectra of the Au/Ti/MoS$_2$/SiO$_2$/Si stack were computed by a normal-incidence transfer-matrix method implemented in Python, with Si treated as a semi-infinite complex substrate. The reflectance $R(\lambda,T)$ was evaluated for each combination of probe wavelength, temperature, and MoS$_2$ thickness. The thermoreflectance coefficient was obtained as \mbox{$C_\mathrm{TR} = R^{-1} \partial R/\partial T$}, with $\partial R/\partial T$ computed by symmetric finite difference over a $\pm1$~K interval centered at the target temperature.

The wavelength dependence was determined by the dispersion of the constituent dielectric functions. Gold was described by the Drude-Lorentz model of Ref.~\cite{Au_model_FarbaniecEakins2023}, which combines a Drude term with a Brendel-Bormann-reduced interband oscillator centered around 2.5~eV. Molybdenum disulfide was described by a two-Lorentz oscillator model representing the combined A/B excitonic weight and the broad C feature. The oscillator parameters were chosen such that the modeled optical constants fall within the range reported for few-layer MoS$_2$ over \mbox{$400$--$700$}~nm~\cite{n-k_MoS2_2019}. This minimal parametrization is sufficient to capture the qualitative thermo-optic response of MoS$_2$ in the probed window. Titanium, silicon dioxide, and silicon were assigned constant optical constants taken from tabulated data at 515~nm~\cite{JohnsonChristy1974,Malitson1965,AspnesStudna1983}. For SiO$_2$ this is an excellent approximation, as its refractive index varies by less than $0.3\%$ across the \mbox{$515$--$600$~nm} probe window~\cite{Malitson1965}. Silicon is noticeably dispersive over the same window~\cite{AspnesStudna1983,Green2008}; however, since the probe field reaching the substrate is strongly attenuated within the Au layer, repeating the simulations with dispersive tabulated silicon data changes the computed reflectance by less than $0.1\%$ and the thermoreflectance coefficient by less than $2\%$ for all MoS$_2$ thicknesses considered.

The temperature dependence enters through linear thermo-optic coefficients applied to the parameters of both Au (Drude damping, plasma energy, interband center and width, oscillator weight, $\epsilon_\infty$) and MoS$_2$ (Lorentz centers and broadenings). The Au coefficients were taken from Ref.~\cite{Au_model_FarbaniecEakins2023}. The MoS$_2$ coefficients were calibrated to reproduce the red-shift and broadening of the A/B and C excitonic features reported between 120 and 300~K~\cite{Ho1998,Liu2020,Golovynskyi2020}. All coefficients were held fixed across temperatures, so that the temperature dependence of $C_\mathrm{TR}$ emerges entirely from the temperature dependence of the underlying dielectric functions. Full parameter values, sources, and equations are provided in the Supporting Information (see Section~\ref{sec:tmm}).

%--------------------------------------------------------------------------------------------------------------------------------------------------------------------------------------------------------------------------------
\medskip
%\newpage
\textbf{Supporting Information} \par %Please delete the Suppporting Information statement if it is not applicable. Please supply Supporting Information in another file. Supporting information should not be provided in .tex format
Supporting Information is available from the Wiley Online Library or from the author.

% Acknowledgements --------------------------------------------------------
\medskip
\textbf{Acknowledgements} \par %delete if not applicable))
The ICN2 is supported by the Severo Ochoa Centres of Excellence programme, Grant CEX2021-001214-S, funded by MCIU/AEI/10.13039.501100011033. KJT acknowledges funding from the European Union’s Horizon 2020 research and innovation program under Grant Agreement No. 101125457 (ERC CoG ‘EQUATE’) and Spanish MCIN/AEI project PID2022-142730NB-I00 ‘HYDROPTO’.
JSR acknowledges funding from MCIN/AEI/10.13039/501100011033 under grant PID2024-162811NB-I00, and the Severo Ochoa Centres of Excellence Program under grant CEX2023-001263-S.

% References --------------------------------------------------------------
\medskip

% Use the following code if you wish to generate your bibliography with BibTeX;
% replace the string "MSP-template" below with the name(s) of
% the BibTeX data base(s) you want to use.
% The resulting bibliography-output (the content of the .bbl file)
% must be pasted back into this file before submission.
% Please also include your BibTeX data base file(s) in your submission
% so that we can re-run BibTeX if necessary.
%
\bibliographystyle{MSP}
\bibliography{references}

\begin{figure}[htpb]
\textbf{Table of Contents}\\
\medskip
  \includegraphics[height=20mm]{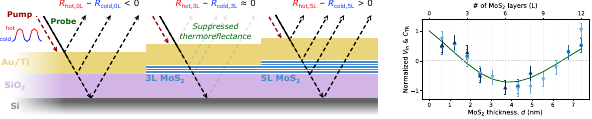}
  \medskip
  \caption*{Controlling the thermoreflectance coefficient of Au-MoS$_2$ transducer metamaterials with the number of MoS$_2$ layers buried underneath a 50~nm gold film. Time-resolved optothermal measurements and transfer-matrix method simulations capture the suppression, sign-reversal and sign-recovered thermoreflectance signal mediated by nanometer-scale thickness changes of the buried MoS$_2$ layer.}
\end{figure}

%%%%%%%%%%%%%%%%%%%%%%%%%%%%%%%%%%%%%%%%%%%%%%%%%%%%%%%%%%
%%%%%%%%%%%%%%%%%%%%%% SUPPORTING INFO %%%%%%%%%%%%%%%%%%%
\newpage

\pagestyle{fancy}
\fancyhf{}                              % clear default header/footer
\fancyhead[L]{\nouppercase{\leftmark}}  % first-level section, right side
\fancyfoot[C]{\thepage}
\renewcommand{\sectionmark}[1]{\markboth{\thesection\ #1}{}}
\rhead{\includegraphics[width=2.5cm]{vch-logo.png}}

\clearpage
\setcounter{figure}{0}
\setcounter{table}{0}
\setcounter{section}{0}

\renewcommand{\thefigure}{S\arabic{figure}}
\renewcommand{\thetable}{S\arabic{table}}
\renewcommand{\thesection}{S\arabic{section}}

\etocdepthtag.toc{supp}

\title{Controlling the Sign and Magnitude of the Thermoreflectance Coefficient of Gold-MoS$_2$ Metamaterials}

\huge Supporting Information for 
\maketitle

\thispagestyle{nosection}

\author{David Saleta Reig*}
\author{Bohai Liu}
\author{Kai Xu}
\author{Juan Sebastián Reparaz}
\author{Klaas-Jan Tielrooij*}

\medskip

\etocsettagdepth{main}{none}
\etocsettagdepth{supp}{subsection}   % or {section} for top-level only
\tableofcontents

\hfill

\section{Sample fabrication and characterization}
\label{sec:SI_SampleFabCharact}

The details of the mechanical exfoliation and dry-transfer approach were reported in Ref.~\cite{S_varghese2021fabrication}. The first step in sample fabrication is preparing the polydimethylsiloxane (PDMS) viscoelastic stamps that we will use to transfer our flakes on the target substrates, followed by a mechanical exfoliation from bulk MoS$_2$. We bring the exfoliation tapes in contact with a PDMS bed previously cut and placed on a glass slide, and subsequently peel the tape producing thin flakes on PDMS ready for inspection and transfer. Prior to transfer we estimate the thickness of our thin flakes ($<$8L) by comparing the discrete variations in optical contrast produced by each MoS$_2$ layer (see Section~\ref{sec:OpticalContrast}). We then transfer our flakes onto SiO$_2$/Si substrates and determine the thickness of thick flakes of interest ($7$--$12$L) using atomic force microscopy (AFM) (see Section~\ref{sec:AFM}). Finally, we use e-beam evaporation to deposit 50~nm (or 100~nm) gold films with 2~nm titanium adhesion layers. With these steps, we fabricated four Au-MoS$_2$ metamaterial samples that collectively cover all thicknesses from 1 to 12L, with multiple regions of each thickness across the sample set. Three of these samples were coated with 50~nm gold, while for the fourth sample we deposited 100~nm gold.

    \subsection{Thickness determination by optical contrast}
    \label{sec:OpticalContrast}

\begin{figure}[htpb]
    \centering
    \includegraphics[width=0.95\linewidth]{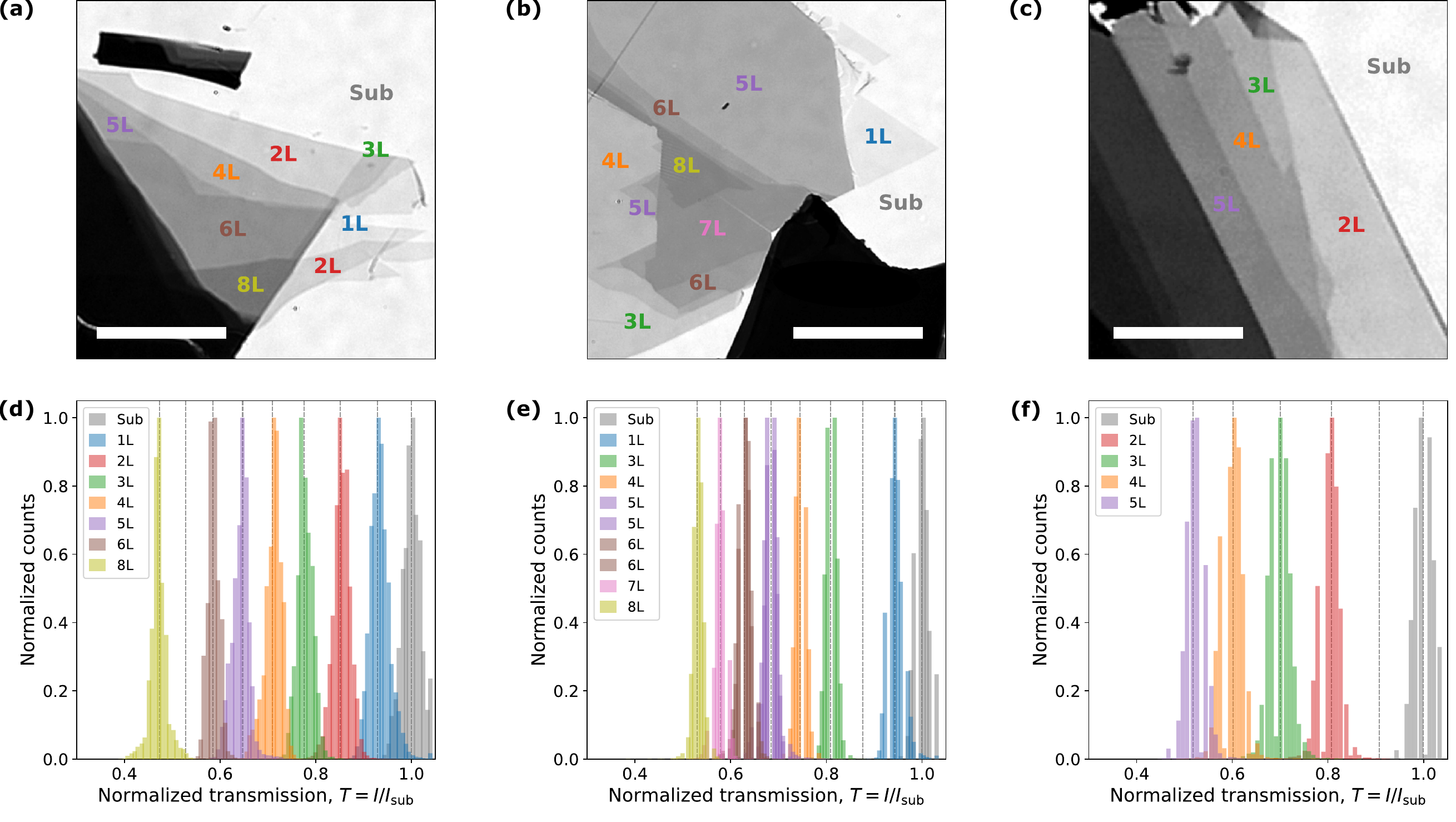}
    \caption{\textbf{Thickness determination by optical contrast.} \textbf{(a--c)}~Green-channel transmission optical images of the exfoliated terrace-like flakes on PDMS prior to transfer onto SiO$_2$. The scale bars are 50~\textmu m for panels (a) and (b), and 20~\textmu m for panel (c). \textbf{(d--e)}~Transmission histograms for each flake region normalized to the intensity of the bare substrate. The thickness of thin MoS$_2$ ($<$8L) is unambiguously assigned by sorting the step-like intensity profile for each layer thickness. The vertical dashed lines serve as a guide to the eye.}
    \label{fig:SI_OpticalContrast}
\end{figure}

    After mechanical exfoliation of bulk MoS$_2$ on PDMS, we identify terrace-like MoS$_2$ flakes under an optical microscope and take transmission optical images. We split the RGB channels and use the green-channel transmission images (see \textbf{Figure~\ref{fig:SI_OpticalContrast}a--c}) to retrieve the transmitted intensity of each region. We plot the region-normalized histograms, the intensity of which we normalize to the substrate's intensity \mbox{$T = I/I_\mathrm{sub}$}. The normalized intensities follow a discrete step-like reduction in transmission upon increasing the number of MoS$_2$ layers. Thanks to the use of terrace-like MoS$_2$ flakes with various thicknesses, we can identify the missing thicknesses and give unambiguous estimates for the number of MoS$_2$ layers of each region with $<$8L (see Figure~\ref{fig:SI_OpticalContrast}\textbf{d--f}). This approach works well for thin TMD flakes because of their large, layer-dependent optical absorption~\cite{taghavi2019thickness}.

\hfill

\newpage
    \subsection{Thickness determination by atomic force microscopy}
    \label{sec:AFM}

\begin{figure}[ht]
    \centering
    \includegraphics[width=0.667\linewidth]{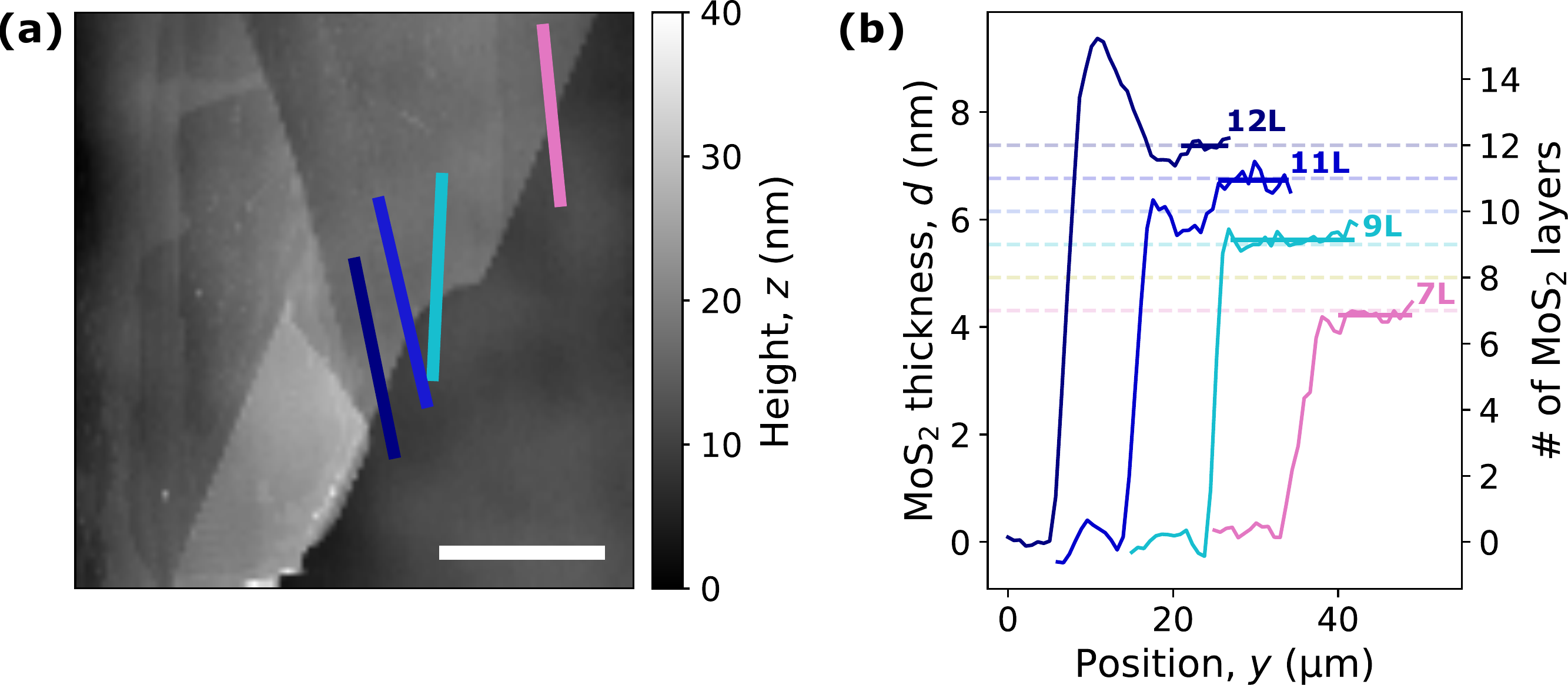}
    \caption{\textbf{Thickness determination by atomic force microscopy.} \textbf{(a)}~Height map of a thick terrace-like MoS$_2$ flake on SiO$_2$/Si substrate. The scalebar is 20~\textmu m. \textbf{(b)}~Linecuts across the solid lines of panel (a) showing the thickness and number of MoS$_2$ layers retrieved from the height difference between MoS$_2$ and substrate.}
    \label{fig:SI_AFM}
\end{figure}

    After transferring the selected terrace-like MoS$_2$ flakes onto SiO$_2$/Si substrates, we perform AFM topography imaging in tapping mode on various regions of interest from our transferred flakes (see \textbf{Figure~\ref{fig:SI_AFM}a}). We retrieve height profiles across the edge of the MoS$_2$ flake for each layer of interest, and measure the height difference between MoS$_2$ and substrate (see Figure~\ref{fig:SI_AFM}\textbf{b}). 

\hfill

\clearpage
\newpage
\section{Time-domain thermoreflectance setup}
\label{sec:Setup}

\begin{figure}[htpb]
    \centering
    \includegraphics[width=\linewidth]{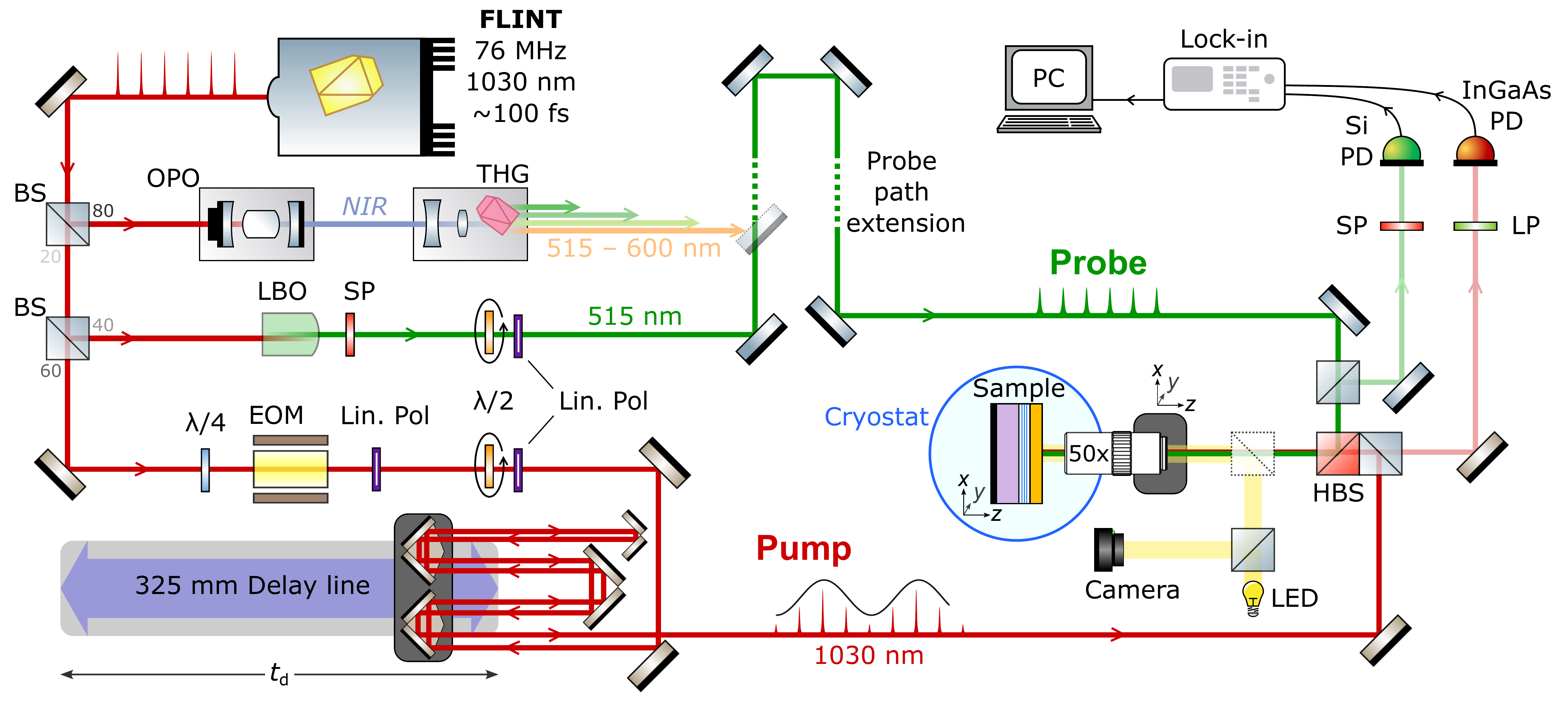}
    \caption{\textbf{Optothermal setup layout.} Schematic of the time-domain thermoreflectance setup used for our optothermal measurements on Au-MoS$_2$ metamaterials. An amplitude-modulated train of pump pulses induces transient reflectivity changes on a sample, which are measured by a train of probe pulses at a variable time delay. OPO: optical parametric oscillator; THG: third-harmonic generation; LBO: lithium triborate crystal; EOM: electro-optic modulator; Lin.~Pol: linear polarizer; $\lambda/4$: quarter-waveplate; $\lambda/2$: rotatable half-waveplate; BS: beamsplitter; HBS: harmonic beamsplitter; LED: light-emitting diode; SP: short-pass filter; LP: long-pass filter; PD: photodiode; $t_\mathrm{d}$: time delay.}
    \label{fig:TDTRsetup}
\end{figure}

We present a scheme of our home-built pump-probe optothermal setup in \textbf{Figure~\ref{fig:TDTRsetup}}. The fundamental laser emits laser pulses (with average power of 14~W and pulse durations of $\sim$100~fs) centered at 1030~nm wavelength, at a repetition rate of 76~MHz. Most power ($\sim$80$\%$) is directed to an optical parametric oscillator (OPO), the signal output of which is fed to a third harmonic conversion box to produce the wavelength-tunable probes (\textit{i.e.}, 515, 525, 530 and 600~nm). The rest of power is split into our standard pump and probe beams. The probe is frequency-doubled (515~nm) using a non-linear lithium triborate crystal. For the pump beam, an electro-optic modulator (EOM) introduces an amplitude modulation in the MHz range, eliminating the need of an external bias voltage thanks to the quarter-waveplate approach. The pump beam passes over a retroreflector-based quadruple-pass delay line which allows us to vary the time delay between pump and probe pulses in the range between $-1$ and $7$~ns. The pump and probe beams recombine at a harmonic beamsplitter, which also allows us to collect the probe beam on a silicon photodiode. A lock-in amplifier demodulates the signal at the modulation frequency of the pump beam. The sample sits in a closed-cycle He cryostat, which we use to vary the temperature from room temperature down to 120~K. 

\hfill

\newpage
\subsection{Laser spot size characterization}

\begin{figure}[htpb]
    \centering
    \includegraphics[width=\linewidth]{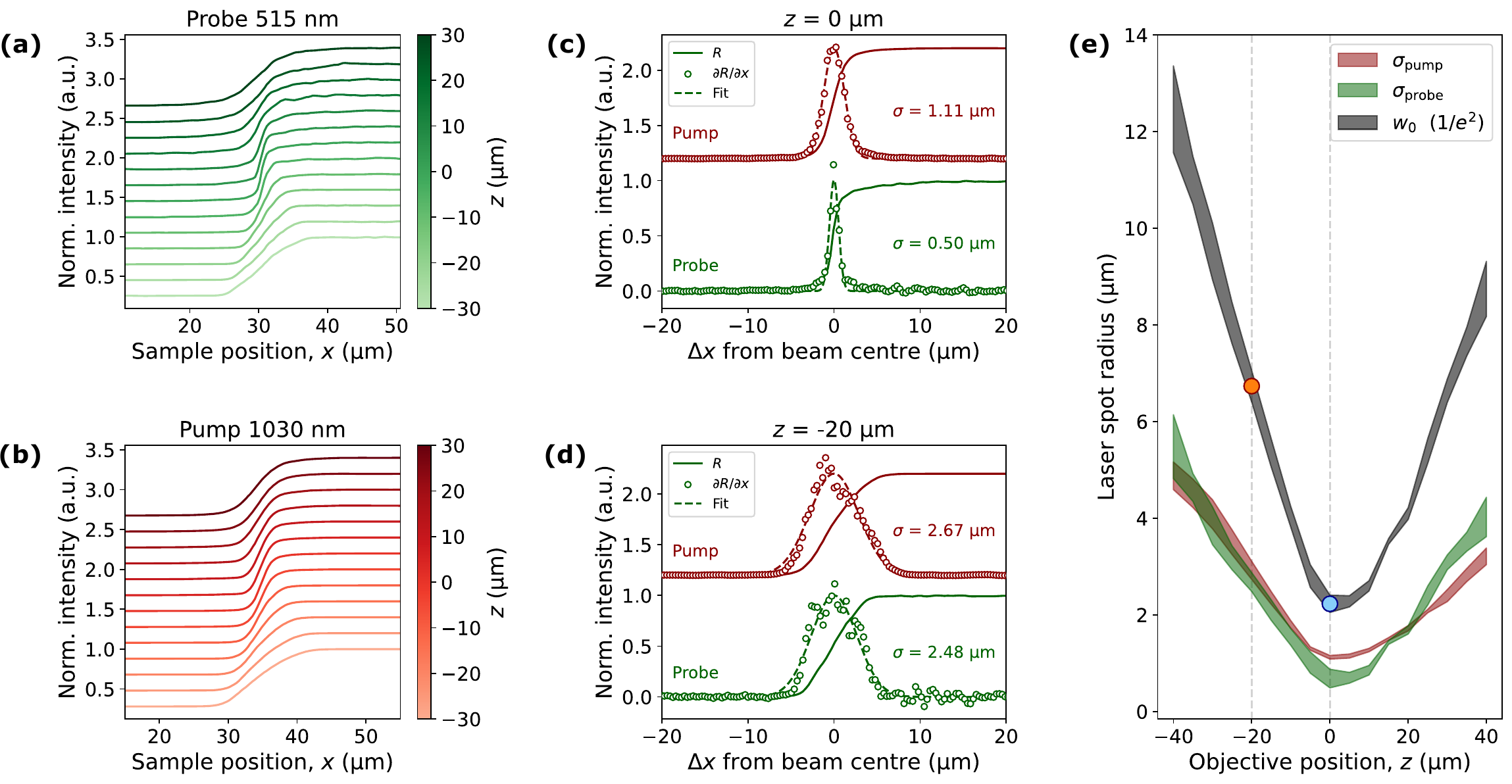}
    \caption{\textbf{Spot size characterization of probe and pump beams by the
    knife-edge method.}
    \textbf{(a,b)}~Normalized reflectivity $R$ of the probe (515~nm) and pump (1030~nm) beams, respectively, recorded while scanning across the edge of a lithographically defined 50~nm-thick gold transducer layer on SiO$_2$. Profiles are collected at multiple objective positions~$z$ (\textit{color-coded}), ranging from far below to above the optical focus at \mbox{$z=0$}. When approaching the focus, the lateral width of the reflectivity step narrows, directly reflecting the reduction in beam spot size.
    \textbf{(c,d)}~Reflectivity profiles (\textit{solid lines}) and their spatial derivatives $\partial R/\partial x$ (\textit{scatters}) for both the probe (\textit{green, lower}) and pump (\textit{red, upper}) beams at the focal plane (\mbox{$z=0$~\textmu m}) and at a defocused position (\mbox{$z=20$~\textmu m}), respectively. Each profile is individually centered at its Gaussian fit position so that the two beams can be compared on the same axis. Gaussian fits to $\partial R/\partial x$ (\textit{dashed lines}) yield the standard deviation~$\sigma$, which is proportional to the beam radius.
    \textbf{(e)}~Fitted $\sigma$ as a function of objective position for the probe (\textit{green}) and pump (\textit{red}) beams, together with the derived RMS average spot radius~$w_0$ using the $1/e^2$ definition (\textit{black}). The shaded areas represent the propagation of errors computed from measurements in $x$ and $y$ directions (to account for ellipticity), and in the case of the pump beam also at the two extreme positions of the delay line (to account for deviations in the beam's collimation). The scatter points highlight the averaged laser spot sizes of 2.2~\textmu m (\textit{blue}) and 6.7~\textmu m (\textit{orange}) used to retrieve the results of Figure~\ref{fig:SI-FIG_RTpos_SignalMagnitude}.}
    \label{fig:knifeEdge}
\end{figure}

We determine the laser spot sizes for pump and probe beams using the knife edge method. Here, we scan the edge of a lithographically defined gold film while collecting the reflectivity of each beam independently. The results show step-like reflectance profiles originating from the reflectivity contrast between SiO$_2$/Si and gold. We repeat these measurements at different heights ($z$) around the laser focus (at \mbox{$z=0$}), resulting in sharper contrast when near the laser focus due to a smaller beam size (see \textbf{Figure~\ref{fig:knifeEdge}a} and~\ref{fig:knifeEdge}\textbf{b} for probe and pump, respectively). We take the spatial derivatives of the step-like reflectance profiles and retrieve the standard widths $\sigma$ from Gaussian fits to the $\partial R/\partial x$ data (see Figure~\ref{fig:knifeEdge}\textbf{c} and~\ref{fig:knifeEdge}\textbf{d}). Finally, we convert our laser radii following the $1/e^2$ definition \mbox{($r=\sqrt{2}\sigma$)} into the RMS average radius $w_0$ (see Figure~\ref{fig:knifeEdge}\textbf{e}):
\begin{equation}
    w_0 = \sqrt{\frac{r^2_\mathrm{pump} + r^2_\mathrm{probe}}{2}}
\end{equation}

We use an RMS average spotsize of $\sim$2~\textmu m for all measurements reported in this work, except for results shown in \textbf{Figure~\ref{fig:SI-FIG_RTpos_SignalMagnitude}b}, where we use an average spot size of $\sim$7~\textmu m and find that the conclusions derived from this work are independent on laser spot size (see Section~\ref{sec:SpotsizeInvariant_CTR_Control}).

\newpage
\subsection{Lock-in amplifier signals}
We overlap our pump and probe beams in space and time, and illuminate the gold surface on our sample. The pump beam, modulated at a frequency of \mbox{$f_\mathrm{mod} = 1.11$~MHz}, introduces dynamic heating inducing reflectivity changes on our sample. We detect the heat-induced reflectivity changes using our unmodulated probe beam. By collecting the probe's reflected light on a photodiode connected to a lock-in amplifier, we demodulate this signal at $f_\mathrm{mod}$ to retrieve the thermoreflectance signal. Finally, we record the thermoreflectance signal as a function of time delay of arrival between pump and probe pulses, thus mapping the temporal dynamics of the sample's optothermal response.

\begin{figure}[htpb]
    \centering
    \includegraphics[width=0.8\linewidth]{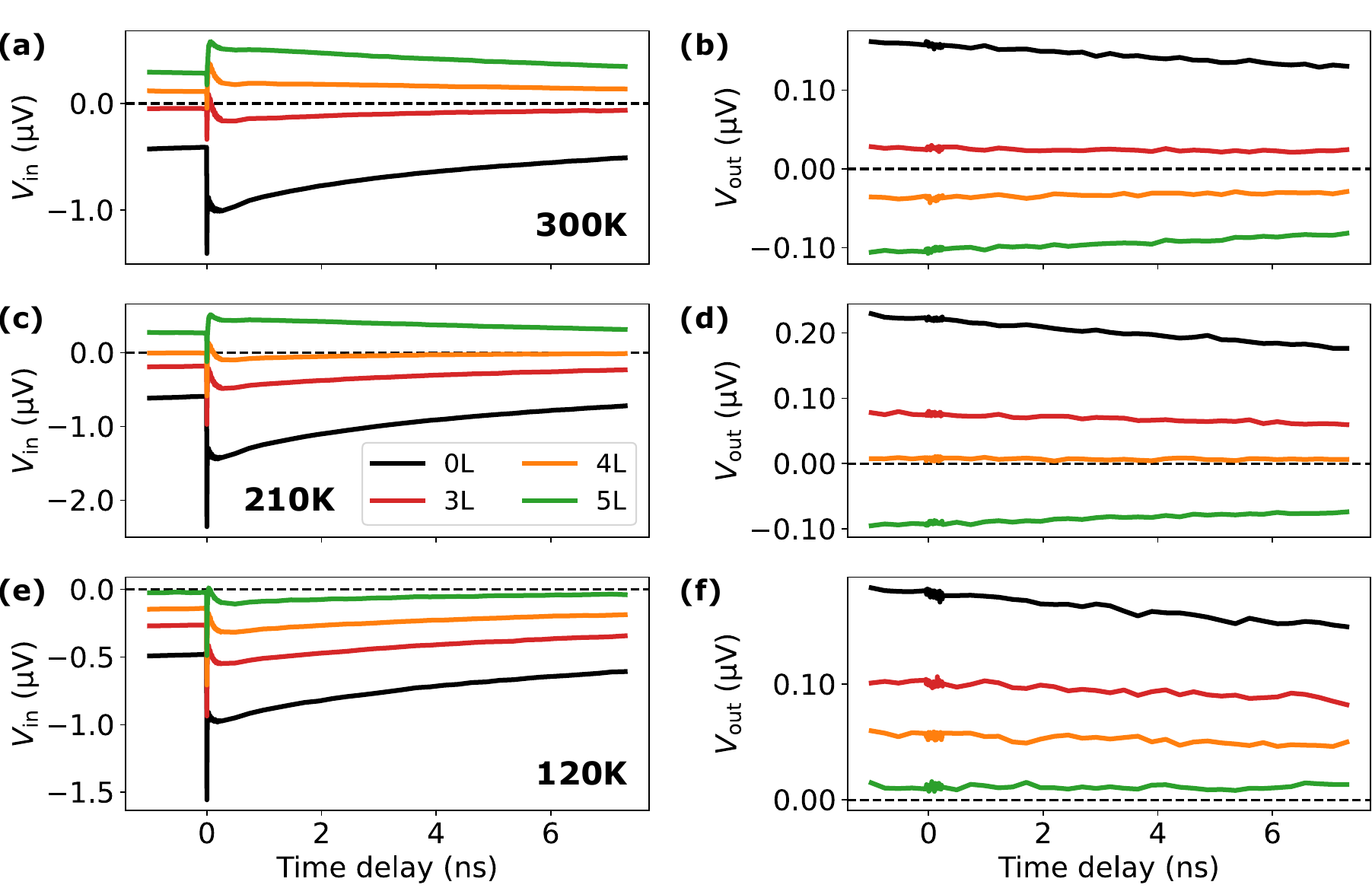}
    \caption{\textbf{Lock-in signals at various ambient temperatures.} \textbf{(a,c,e)}~In-phase and \textbf{(b,d,f)}~out-of-phase lock-in signals for 0L (\textit{black}), 3L (\textit{red}), 4L (\textit{orange}) and 5L (\textit{green}) MoS$_2$ regions at (a,b)~300~K, (c,d)~210~K, and (e,f)~120~K. The 3L region shows minimum signal magnitude at 300~K, the 4L region at 210~K and the 5L region at 120~K. This trend is independent of time delay of arrival between pump and probe pulses, and is consistent in both the $V_\mathrm{in}$ and the $V_\mathrm{out}$ signals. }
    \label{fig:MinimumSignals_vs_T}
\end{figure}

The lock-in amplifier produces in-phase $V_\mathrm{in}$ and out-of-phase $V_\mathrm{out}$ signal components. The in-phase signal contains information of pulsed heat dissipation, while the out-of-phase component relates to the modulated heat dissipation given by the hot and cold pump cycles. Following the standard TDTR data analysis~\cite{S_Cahill2004,S_jiang2018tutorial}, we correct for the phase offset introduced by the instrumentation prior to analyzing thermoreflectance signal magnitude. For this, we perform rotation of the signal in the complex plane ensuring smooth $V_out$ signals across negative to positive time delays. We show exemplary $V_\mathrm{in}$ and $V_\mathrm{out}$ signals from optothermal measurements as a function of ambient temperature for 0, 3, 4, 5L MoS$_2$ in \textbf{Figure~\ref{fig:MinimumSignals_vs_T}}. Our complex-plane-rotation routine individually finds the best phase for each dataset, and the retrieved instrumentation phase offset typically varies within less than $\pm 1$\textdegree~between measurement (with MoS$_2$, $n$L) and control (without MoS$_2$, 0L).

\hfill

%\clearpage
\newpage
\section{Transfer-matrix method simulations}
\label{sec:tmm}

The optical response of the sample was modeled with the transfer-matrix method (TMM), an exact solution of Maxwell's equations for a stack of planar, homogeneous, isotropic layers. The simulated structure reproduces the fabricated samples and consists, from the incidence side, of a semi-infinite air ambient, a $50$~nm gold film, a $2$~nm titanium adhesion layer, the MoS$_2$ flake whose thickness is varied between $0$ and $12$L with a layer thickness of 6.15~\AA/L, a $285$~nm SiO$_2$ spacer, and a semi-infinite silicon substrate. Each layer~$j$ is characterized by a complex refractive index
\mbox{$\tilde n_j(\lambda,T) = n_j + i k_j = \sqrt{\varepsilon_j(\lambda,T)}$} and a thickness~$d_j$; photon energy and wavelength $\lambda$ are related by \mbox{$E\,[\mathrm{eV}] \simeq 1239.84/\lambda\,[\mathrm{nm}]$}. The complete stack is summarized in \textbf{Table~\ref{tab:stack}}.

\begin{figure}[htpb]
    \centering
    \includegraphics[width=0.667\linewidth]{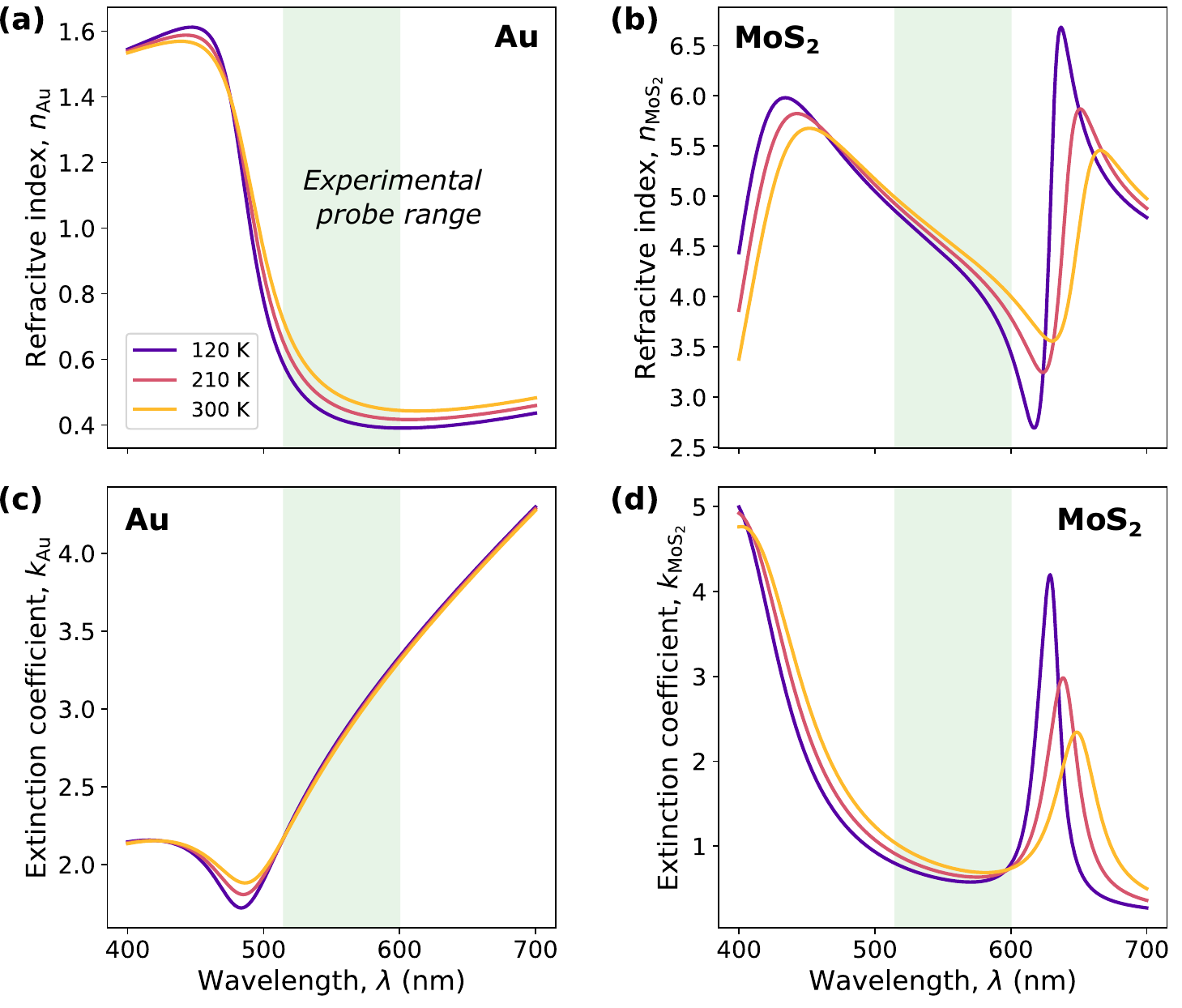}
    \caption{\textbf{Dielectric models for Au and MoS$_2$.} Wavelength-dependent \textbf{(a,b)}~refractive indices $n$ and \textbf{(c,d)}~extinction coefficients $k$ at 300 (\textit{yellow}), 210 (\textit{magenta}) and 120~K (\textit{purple}); for \textbf{(a,c)}~Au, retrieved from the model reported in Ref.~\cite{S_Au_model_FarbaniecEakins2023}, and for \textbf{(b,d)}~MoS$_2$, built from a two-Lorentz oscillator model. The shaded green area represents the experimental probe wavelength range. }
    \label{fig:SI-fig_dielectrics}
\end{figure}

All experiments and TMM simulations reported here are performed at normal incidence such that within each layer the field accumulates a complex phase thickness
\begin{equation}
  \delta_j = \frac{2\pi}{\lambda}\,d_j\,\tilde n_j  ,%\cos\theta_j ,
\end{equation}
and the characteristic \mbox{$2\times2$} matrices of all layers, together with the Fresnel coefficients of each interface, are multiplied to yield the transfer matrix of the full stack. The reflectance~$R$, transmittance~$T$ and absorbance \mbox{$A = 1 - R - T$} are then obtained from its elements. The layer thicknesses are held fixed (thermal expansion is neglected over the experimental range), so that the temperature dependence of the simulated spectra enters exclusively through the dielectric functions of gold and MoS$_2$ described below (see \textbf{Figure~\ref{fig:SI-fig_dielectrics}}). The titanium adhesion layer, the SiO$_2$ spacer, and the silicon substrate are assigned constant complex refractive indices, taken as wavelength- and temperature-independent over the spectral and temperature range probed here; their values are listed in Table~\ref{tab:stack}.

% ----------------------------------------------------------------------------
%  TABLE 1 — layer stack
% ----------------------------------------------------------------------------
\hfill
\begin{table}[htpb]
  \centering
  \caption{\textbf{Material stack used in the transfer-matrix simulations.} The structure is bounded above by air (\mbox{$n=1$}, \mbox{$k=0$}) and terminated below by a semi-infinite silicon substrate. Titanium, SiO$_2$ and silicon are treated as non-dispersive and assigned constant refractive indices \mbox{$\tilde n = n + ik$}. \textsuperscript{*}The values quoted for gold and MoS$_2$ are representative, evaluated at \mbox{$\lambda = 515$~nm} and \mbox{$T_{\mathrm{ref}} = 300$~K}; the dielectric functions of both materials are wavelength- and temperature-dependent (see Figure~\ref{fig:SI-fig_dielectrics}), described by the dispersive models of Section~\ref{sec:tmm} (model parameters in \textbf{Table~\ref{tab:params}}).}
  \label{tab:stack}
  \begin{tabular}{@{}ccc@{\hspace{2em}}D{.}{.}{2}@{\hspace{2.5em}}D{.}{.}{2}@{\hspace{0.5em}}}
    \toprule
    Layer & Material & Thickness (nm) & \multicolumn{1}{l}{$n$} & \multicolumn{1}{l}{$k$} \\
    \midrule
    1 & Au      & 50       & 0.71\textsuperscript{*} & 2.18\textsuperscript{*} \\
    2 & Ti      & 2        & 1.80                    & 1.98                    \\
    3 & MoS$_2$ & 0--8     & 4.98\textsuperscript{*} & 1.04\textsuperscript{*} \\
    4 & SiO$_2$ & 285      & 1.46                    & 0.00                    \\
    5 & Si      & $\infty$ & 4.22                    & 0.07                    \\
    \bottomrule
  \end{tabular}
\end{table}

\begin{figure}[htbp]
    \centering
    \includegraphics[width=0.333\linewidth]{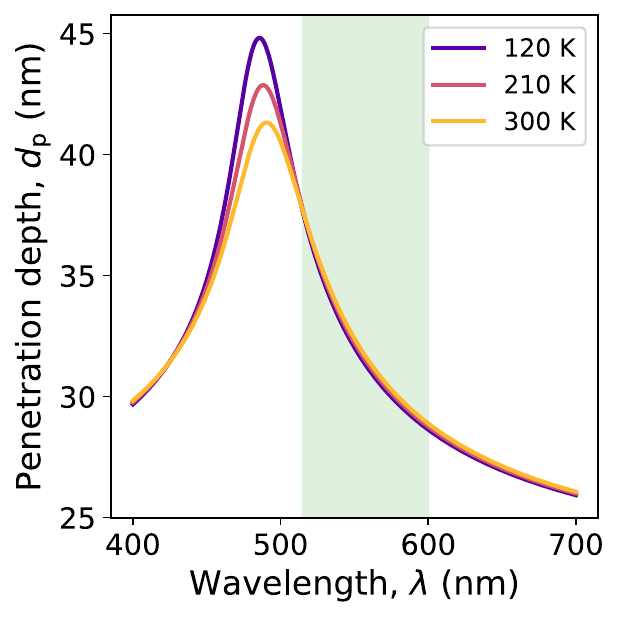}
    \caption{\textbf{Simulated optical penetration depth of gold}. Wavelength-dependent optical penetration depth \mbox{$d_p = \lambda/(2\pi k)$} retrieved using the gold dielectric model from Ref.~\cite{S_Au_model_FarbaniecEakins2023} and the gold parameters in Table~\ref{tab:params}. The shaded green area represents the experimental probe wavelength range.}
    \label{fig:penetrationDepth}
\end{figure}

We calculate the wavelength-dependent refractive indices $n$ and extinction coefficients $k$ of gold and MoS$_2$ (see Figure~\ref{fig:SI-fig_dielectrics}), and optical penetration depth of gold (see \textbf{Figure~\ref{fig:penetrationDepth}}) using the dielectric functions of gold (see Section~\ref{sec:tmm-au}) and MoS$_2$ (see Section~\ref{sec:tmm-mos2}).

\newpage
\subsection{Gold dielectric function}
\label{sec:tmm-au}
 
The dielectric function of the gold film follows the model of Farbaniec and Eakins~\cite{S_Au_model_FarbaniecEakins2023}, in which a Drude term is combined with a single reduced Brendel-Bormann interband oscillator,
\begin{equation}
  \varepsilon_{\mathrm{Au}}(E,T) = \varepsilon_\infty(T)
  + \varepsilon_{\mathrm{D}}(E,T) + \varepsilon_{\mathrm{BB}}(E,T) .
\end{equation}
The intraband (Drude) contribution is
\begin{equation}
  \varepsilon_{\mathrm{D}}(E,T) = -\,\frac{E_P^{2}}{E\,(E + i\gamma)} ,
\end{equation}
with $E_P$ an effective plasma energy and $\gamma$ the Drude damping. The interband (Brendel-Bormann) contribution is written in closed form as
\begin{equation}
  \varepsilon_{\mathrm{BB}}(E,T) =
  \frac{w_B\,E_P^{2}}{2\,a}\,
  \ln\!\left(\frac{E_B + a}{E_B - a}\right),
  \qquad
  a = \sqrt{E\,(E + i\Gamma)} ,
\end{equation}
where $w_B$ is the interband oscillator strength, $E_B$ the interband (critical-point) transition energy and $\Gamma$ the interband broadening. In the model of Ref.~\cite{S_Au_model_FarbaniecEakins2023}, $E_P$ is treated as an effective fit parameter rather than the true plasma energy of gold. We retain this parametrization but adjust $E_P$ from their fitted value of $11.81$~eV to $9.00$~eV, bringing it closer to the plasma energy of bulk gold ($\approx 8.45$~eV) while keeping it within the range commonly reported for thin gold films~\cite{OpticalUltrathin_Au_2023}.

\hfill
\subsection{\texorpdfstring{MoS$_2$ dielectric function}{MoS2 dielectric function}}
\label{sec:tmm-mos2}
 
The MoS$_2$ layer is described by a minimal phenomenological excitonic model comprising a background permittivity and two Lorentz oscillators,
\begin{equation}
  \varepsilon_{\mathrm{MoS_2}}(E,T) = \varepsilon_\infty
  + \sum_{j\,\in\,\{\mathrm{A+B},\,\mathrm{C}\}}
  \frac{A_j\,E_{0,j}^{2}(T)}{E_{0,j}^{2}(T) - E^{2} - i\,\Gamma_j(T)\,E} .
\end{equation}
The first oscillator (A$+$B) combines the A and B excitonic resonances into a single effective feature, while the second oscillator (C) represents the broad, strong C feature. The oscillator strengths $A_j$, center energies $E_{0,j}$ and broadenings $\Gamma_j$ were tuned to reproduce physically reasonable optical constants in the $500$--$650$~nm window at room temperature (\mbox{$n\approx4-5$}, \mbox{$k\approx1-1.5$}) and to yield a sizable optothermal response on the high-energy flank of the B feature and within the C feature, the spectral regime probed in this work.

\newpage
\subsection{Temperature dependence}
\label{sec:tmm-temperature}
 
All temperature-dependent quantities are described by a single linear law about the reference temperature $T_{\mathrm{ref}} = 300$~K. Every such parameter $p$ of both dielectric models obeys
\begin{equation}
  p(T) = p(T_{\mathrm{ref}})
  + \frac{\mathrm{d}p}{\mathrm{d}T}\,\bigl(T - T_{\mathrm{ref}}\bigr).
  \label{eq:tdep}
\end{equation}
This applies to $\{\varepsilon_\infty, E_P, \gamma, w_B, E_B, \Gamma\}$ for gold and to the oscillator centers $E_{0,j}$ and broadenings $\Gamma_j$ for MoS$_2$. The MoS$_2$ oscillator strengths $A_j$ and background permittivity are assumed temperature-independent, as are the optical constants of Ti, SiO$_2$ and Si. With the sign convention of Equation~\eqref{eq:tdep}, a negative $\mathrm{d}E_{0}/\mathrm{d}T$ corresponds to the usual excitonic red-shift (blue-shift) on heating (cooling), while a positive $\mathrm{d}\Gamma/\mathrm{d}T$ describes thermal broadening of the resonance. The complete parameter set is presented in Table~\ref{tab:params}.

% ----------------------------------------------------------------------------
%  TABLE 2 — dispersive model parameters
% ----------------------------------------------------------------------------
\hfill
\begin{table}[htpb]
  \centering
  \caption{\textbf{Parameters of the dispersive dielectric models of gold and MoS$_2$}, evaluated at the reference temperature \mbox{$T_{\mathrm{ref}} = 300$~K}. Each parameter $p$ varies linearly with temperature following Equation~\eqref{eq:tdep}; the tabulated temperature coefficient is the slope $\mathrm{d}p/\mathrm{d}T$ scaled by a factor $10^{4}$, and a dash marks the temperature-independent parameters. For gold, the parameters adjusted with respect to the values used in Ref.~\cite{S_Au_model_FarbaniecEakins2023} are preceded by the original value in parentheses. The two MoS$_2$ oscillators, labeled A$+$B and C, represent the merged A/B excitons and the higher-energy C feature, respectively.}
  \label{tab:params}
  \begin{tabular}{@{}>{\centering\arraybackslash}p{2.7cm}crrrr@{}}
    \toprule
    \multirow{2}{2.7cm}{\centering Material\,/\,model} & \multirow{2}{*}{Symbol}
      & \multicolumn{2}{c}{At $T_{\mathrm{ref}} = 300$~K}
      & \multicolumn{2}{c}{Temperature coefficient} \\
    \cmidrule(lr){3-4}\cmidrule(lr){5-6}
      & & Value & Unit & $\mathrm{d}p/\mathrm{d}T\times10^{4}$ & Unit \\
    \midrule
    \multirow{6}{2.7cm}{\centering\textbf{Gold}~\cite{S_Au_model_FarbaniecEakins2023}\\[4pt]{\footnotesize Drude\\[2pt] (intraband)\\[2pt]$+$\\[2pt]Brendel-Bormann\\(interband)}}
      & $\varepsilon_\infty$ & $1.03$           &               & $6.03$           & $\mathrm{K^{-1}}$     \\
      & $E_P$                & $(11.81)\;9.00$  & $\mathrm{eV}$ & $0.48$           & $\mathrm{eV\,K^{-1}}$ \\
      & $\gamma$             & $0.27$           & $\mathrm{eV}$ & $(1.77)\;1.20$   & $\mathrm{eV\,K^{-1}}$ \\
      & $w_B$                & $0.16$           &               & $-0.01$          & $\mathrm{K^{-1}}$     \\
      & $E_B$                & $(2.51)\;2.57$   & $\mathrm{eV}$ & $(-0.85)$\;$-1.50$ & $\mathrm{eV\,K^{-1}}$ \\
      & $\Gamma$             & $0.29$           & $\mathrm{eV}$ & $3.75$           & $\mathrm{eV\,K^{-1}}$ \\
    \midrule
    \multirow{7}{2.7cm}{\centering\textbf{MoS$_2$}\\[4pt]{\footnotesize background permittivity\\[2pt]$+$\\[2pt]two Lorentz\\(excitonic) oscillators}}
      & $\varepsilon_\infty$    & $5.00$ &               & ---     &                       \\
      & $A_{\mathrm{A+B}}$      & $1.00$ &               & ---     &                       \\
      & $E_{0,\mathrm{A+B}}$    & $1.90$ & $\mathrm{eV}$ & $-3.50$ & $\mathrm{eV\,K^{-1}}$ \\
      & $\Gamma_{\mathrm{A+B}}$ & $0.10$ & $\mathrm{eV}$ & $3.00$  & $\mathrm{eV\,K^{-1}}$ \\
      & $A_{\mathrm{C}}$        & $8.50$ &               & ---     &                       \\
      & $E_{0,\mathrm{C}}$      & $2.95$ & $\mathrm{eV}$ & $-5.00$ & $\mathrm{eV\,K^{-1}}$ \\
      & $\Gamma_{\mathrm{C}}$   & $0.60$ & $\mathrm{eV}$ & $3.00$  & $\mathrm{eV\,K^{-1}}$ \\
    \bottomrule
  \end{tabular}
\end{table}

\clearpage
\newpage
\section{Experimental and simulated reflectance}

\begin{figure}[htpb]
    \centering
    \includegraphics[width=\linewidth]{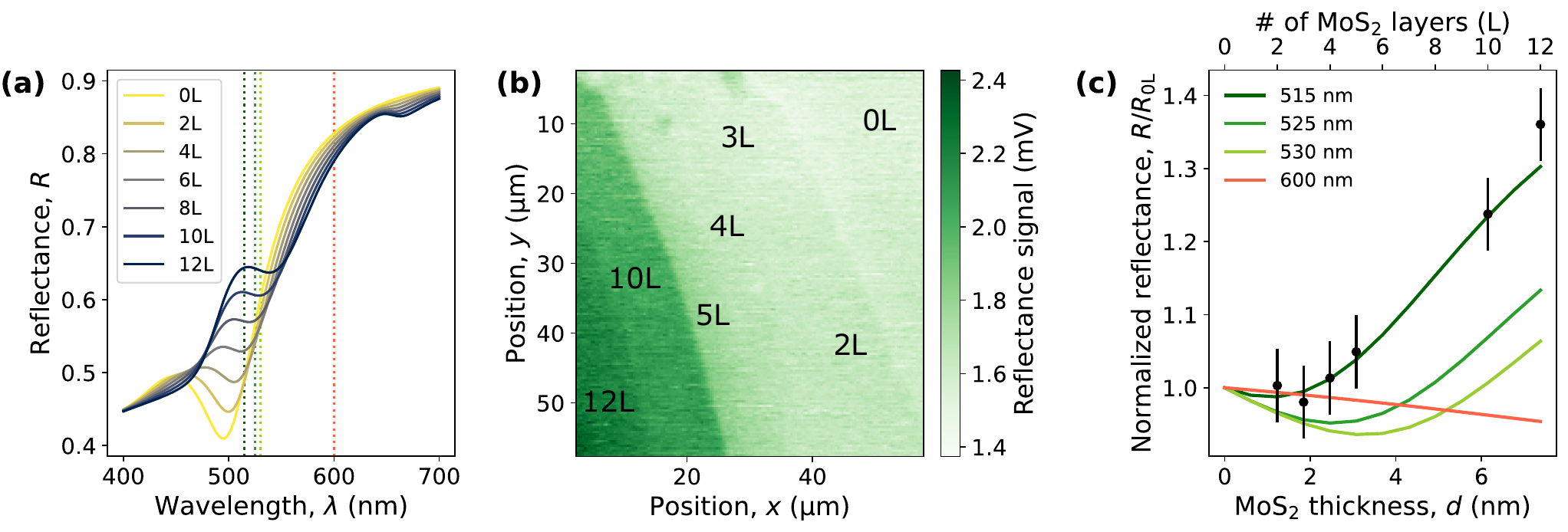}
    \caption{\textbf{Reflectance simulations and experiments.} \textbf{(a)}~Transfer-matrix method simulations of the reflectance in a Au/Ti/MoS$_2$/SiO$_2$/Si stack when varying the number of MoS$_2$ layers. \textbf{(b)}~Probe reflectance map obtained with the 515~nm probe beam, using a mechanical chopper at 240~Hz and lock-in detection. For optical and thermoreflectance images of this sample region see Figure~\ref{fig:RT_thermoreflectance_Figure1}\textbf{b} and~\ref{fig:RT_thermoreflectance_Figure1}\textbf{c}.  \textbf{(c)}~Normalized reflectance ($R/R_\mathrm{0L}$) simulated at different wavelengths (\textit{lines}), with experimental values at 515~nm (\textit{scatters}) extracted from panel (b).}
    \label{fig:SI_fig_Reflectivity}
\end{figure}

To check if the sign-changing thermoreflectance signal is dominated by the reflectivity changes induced by the presence of MoS$_2$ we simulate the wavelength-dependent reflectance of our experimental sample stack (Au/Ti/MoS$_2$/SiO$_2$/Si), producing similar results to those reported for bulk Au in Ref.~\cite{S_Au_model_FarbaniecEakins2023}. At 515~nm wavelengths, the presence of MoS$_2$ induces significant reflectivity differences, with MoS$_2$ increasing the reflectance only when its thickness is thicker than a few layers ($>$4L). Reflectivity differences alone cannot explain the sign-changing thermoreflectance trend obtained for Au-MoS$_2$ metamaterials with varying MoS$_2$ thickness (see \textbf{Figure~\ref{fig:SI_fig_Reflectivity}}).

While MoS$_2$ strongly absorbs light at green wavelengths (\textit{i.e.}~the spectral flank of the B-C excitons), the reflectance $R$ from the sample at 515~nm experimentally changes by less than $\sim$5$\%$ when introducing MoS$_2$ flakes with $<$5L thickness (see Figure~\ref{fig:SI_fig_Reflectivity}b). This is in strong contrast with the thermoreflectance signal $\partial R/\partial T$, which experiences a sign-reversal at these thicknesses. 

As MoS$_2$ is transparent at our pump wavelength of 1030~nm, the effects of pump beam absorption can be neglected. Our TMM simulations confirm that, because our gold films have a thickness of 50~nm (\textit{i.e.}~approaching the optical penetration depth in gold at visible wavelengths), the electric field of the probe light reaches the buried MoS$_2$ flake and the various reflections from the different interfaces are prone to interference, affecting the total reflectance from the sample and therefore its temperature derivative. 

\hfill

\newpage
\section{The influence of gold thickness}
\label{sec:AuThickness100nm}
\begin{figure}[htbp]
    \centering
    \includegraphics[width=\linewidth]{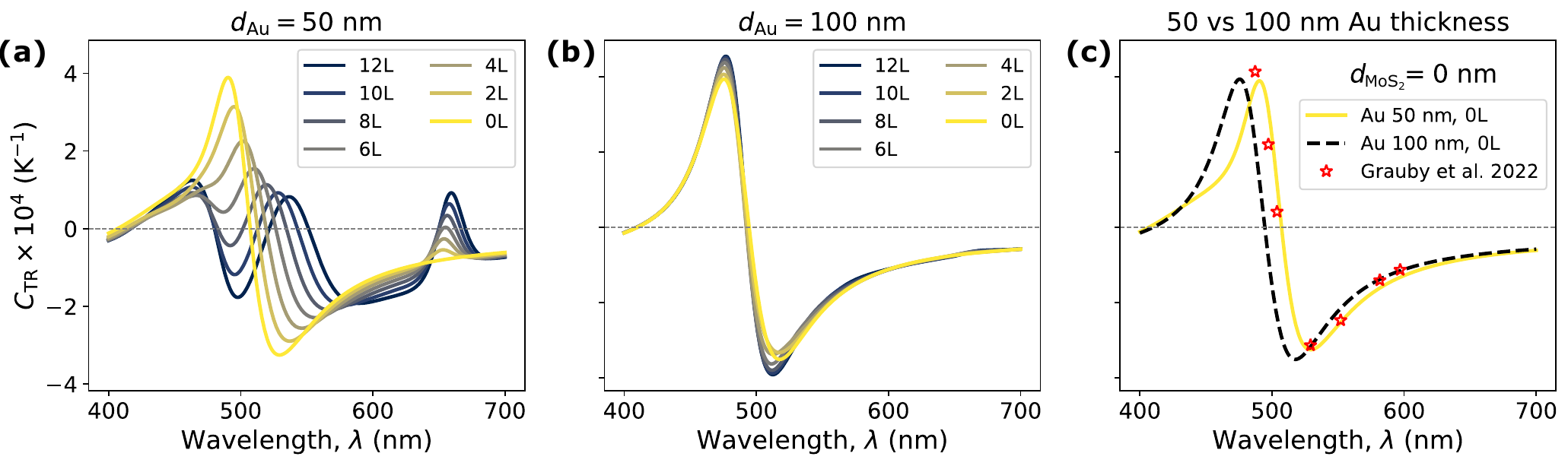}
    \caption{\textbf{The role of Au thickness.} \textbf{(a)}~Transfer-matrix method simulations on a Au/Ti/MoS$_2$/SiO$_2$/Si sample geometry with MoS$_2$ thickness varied from 0 to 12L. The thermoreflectance is strongly influenced by the MoS$_2$ layers underneath the optothermally thin (50~nm) gold film. \textbf{(b)}~When using 100~nm gold, the film is optothermally thick and the influence of the MoS$_2$ underlayers on the effective thermoreflectance becomes vanishingly small. \textbf{(c)}~Comparison between 50 and 100~nm Au thickness, together with results from Ref.~\cite{S_CTR_howMeasureTeTph_2022} on a 50~nm gold film. For 50~nm thin gold, the zero-crossing red-shifts with respect to that of 100~nm Au, which results in probing closer to the zero-crossing point when using 515~nm and, thus, in a stronger influence on the effective thermoreflectance signal upon introducing few-nm thick MoS$_2$ underneath.}
    \label{fig:AuThickness_Sim}
\end{figure}

% - Fig: 100nm gold transducers vs thickness (show Vin, Vout) 
% - Simulations 50 vs 100nm gold... DONE
For bulk gold, we would expect a maximum thermoreflectance signal at $515$~nm based on the literature-reported thermoreflectance spectra~\cite{S_CTR_vs_T_metals_2015, S_Au_model_FarbaniecEakins2023}. 
To further understand our observations, we study the influence of gold thickness simulating the same stacks using 50~nm (see \textbf{Figure~\ref{fig:AuThickness_Sim}a}) and 100~nm (see Figure~\ref{fig:AuThickness_Sim}\textbf{b}) gold thickness. 
First, our simulations reveal a red-shifted thermoreflectance spectrum for 50~nm gold films as compared to thicker films (see Figure~\ref{fig:AuThickness_Sim}\textbf{c}), with 100~nm films already approaching the behavior of bulk gold. This red-shift in the $C_\mathrm{TR}$ spectra when thinning gold down to 50~nm brings the zero-crossing point closer to our experimental probe wavelength of $515$~nm (\textit{i.e.},~\mbox{$C_\mathrm{TR} = 0$} at \mbox{$\lambda=507$~nm}), and the maximum thermoreflectance coefficient now sits at $\sim$525~nm. Our results for Au/Ti/SiO$_2$/Si using 50~nm gold thickness agree well with those reported in Ref.~\cite{S_CTR_howMeasureTeTph_2022}, and using 100~nm gold thickness agree with those reported for bulk-like gold in Ref.~\cite{S_Au_model_FarbaniecEakins2023}. Second, our simulations confirm that the presence of MoS$_2$ significantly affects the thermoreflectance of the stack only when gold is optothermally thin (\mbox{$d_p = 35$--$40$~nm}). 

\hfill

% - Other terraces, TDTR maps (3 in total) and traces (similar as Fig1)
\newpage
\section{\texorpdfstring{Robust thermoreflectance engineering in Au-MoS$_2$ metamaterials}{Robust thermoreflectance engineering in Au/MoS2 metamaterials}}
\label{sec:robust_TR}

Here, we explore the reproducibility and consistency of our observations (independently of the lock-in signal being analyzed, see Section~\ref{sec:VinVout_sametrend}) against the influence of: ($i$)~laser spot sizes (see Section~\ref{sec:SpotsizeInvariant_CTR_Control}); ($ii$)~sample-to-sample variations (see Section~\ref{sec:sampletosampleVariations}); ($iii$)~sample imperfections (see Section~\ref{sec:Cracks}); ($iv$)~modulation frequency of the pump beam (see Section~\ref{sec:fmodInvariantControl}); and ($v$)~the technique employed (see Section~\ref{sec:FDTR}). We find none of these to affect our conclusion that it is possible to engineer the thermoreflectance coefficient of a Au-MoS$_2$ metamaterial by controlling the number of MoS$_2$ layers buried underneath a 50~nm thick gold film.

\subsection{Lock-in-signal-independent observations}
\label{sec:VinVout_sametrend}

\begin{figure}[htpb]
    \centering
    \includegraphics[width=0.7\linewidth]{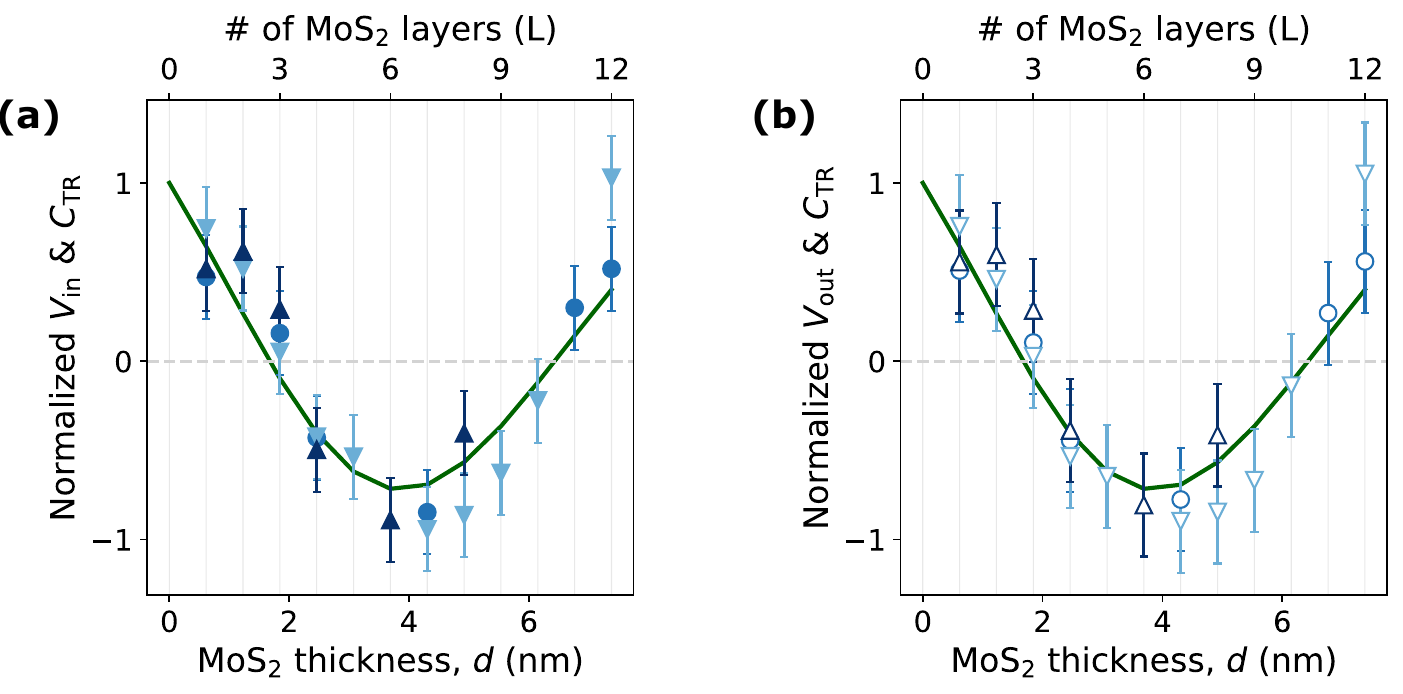}
    \caption{\textbf{Lock-in-signal-independent observations.} Normalized \textbf{(a)}~in-phase $V_\mathrm{in}$ and \textbf{(b)}~out-of-phase lock-in signals obtained as a function of the number of MoS$_2$ layers (\textit{scatters}) and TMM-simulated thermoreflectance coefficient $C_\mathrm{TR}$ (\textit{green solid line}). }
    \label{fig:VinVout_sametrend}
\end{figure}

Throughout the manuscript we have shown the in-phase lock-in signals $V_\mathrm{in}$ and compared them to our simulated thermoreflectance coefficient $C_\mathrm{TR}$. Nevertheless, our observations also apply to the out-of-phase lock-in signals $V_\mathrm{out}$. The normalized signal magnitude for both $V_\mathrm{in}$ and $V_\mathrm{out}$ scales in the exact same way with MoS$_2$ flake thickness (see \textbf{Figure~\ref{fig:VinVout_sametrend}}). This is expected, as both these signals have the same dependencies on the optical and optothermal properties of the transducer (meta)material, with the signal magnitude being influenced by the reflectance and thermoreflectance coefficients. Because the probe's reflectivity is not strongly affected when introducing thin (1--5 layers) MoS$_2$ flakes underneath the 50~nm gold film, we attribute our findings to a modified thermoreflectance coefficient mediated by the buried MoS$_2$. Thanks to our sub-nm control of 2D flake thickness, we exploit the interference of incident and reflected optothermal fields at sub-wavelength scales to engineer the effective $C_\mathrm{TR}$ of Au-MoS$_2$ metamaterials.

\hfill

\newpage
\subsection{Spot-size-invariant thermoreflectance control}
\label{sec:SpotsizeInvariant_CTR_Control}

\begin{figure}[htpb]
    \centering
    \includegraphics[width=0.7\linewidth]{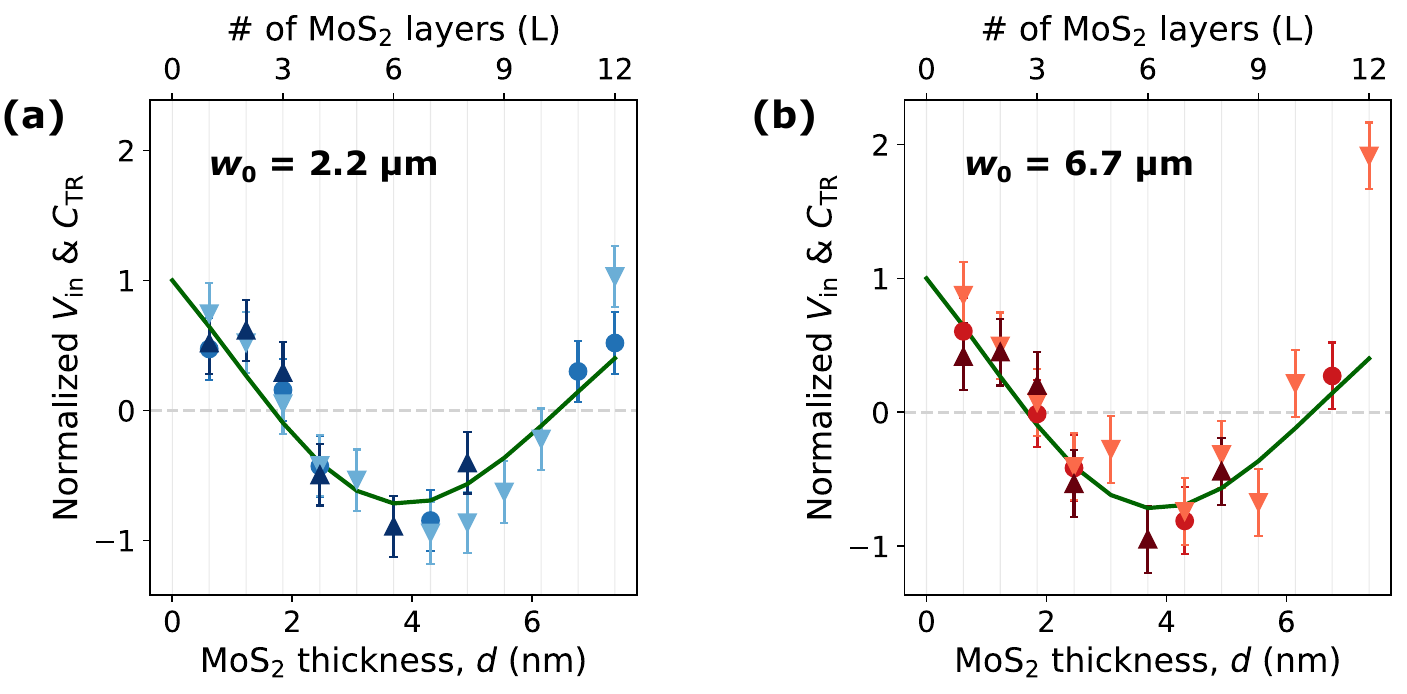}
    \caption{\textbf{Spot-size-independent thermoreflectance control.} Experimental results (\textit{scatters}) and simulations (\textit{lines}) at \mbox{$T=300$~K} and \mbox{$\lambda_\mathrm{pr} = 515$~nm} using \textbf{(a)}~tight focused beams with RMS average laser spot size \mbox{$w_0 \approx 2$~\textmu m}, \textit{i.e.}~the data presented in Figure~\ref{fig:CTRvsWvl_Figure2}\textbf{a}; and \textbf{(b)}~using slightly defocused beams with \mbox{$w_0 \approx 6$~\textmu m}.}
    \label{fig:SI-FIG_RTpos_SignalMagnitude}
\end{figure}

All measurements were performed at best-focusing conditions, using an RMS average spotsize \mbox{$w_0 = 2.2$~\textmu m}. We repeated most measurements using slightly defocused beams, with an RMS average spotsize three times larger than that used in the main text. Using even larger beam sizes is challenging here due to the finite lateral sizes of our flake regions. Figure~\ref{fig:SI-FIG_RTpos_SignalMagnitude}\textbf{b} presents measurements performed with a laser spot size of \mbox{$w_0 = 6.7$~\textmu m}. The thermoreflectance signal retains the same scaling trend with MoS$_2$ thickness, indicating that the observed behavior is independent of the laser spot size.

\hfill

\newpage
\subsection{Modulation-frequency-invariant thermoreflectance control}
\label{sec:fmodInvariantControl}
\begin{figure}[htpb]
    \centering
    \includegraphics[width=0.7\linewidth]{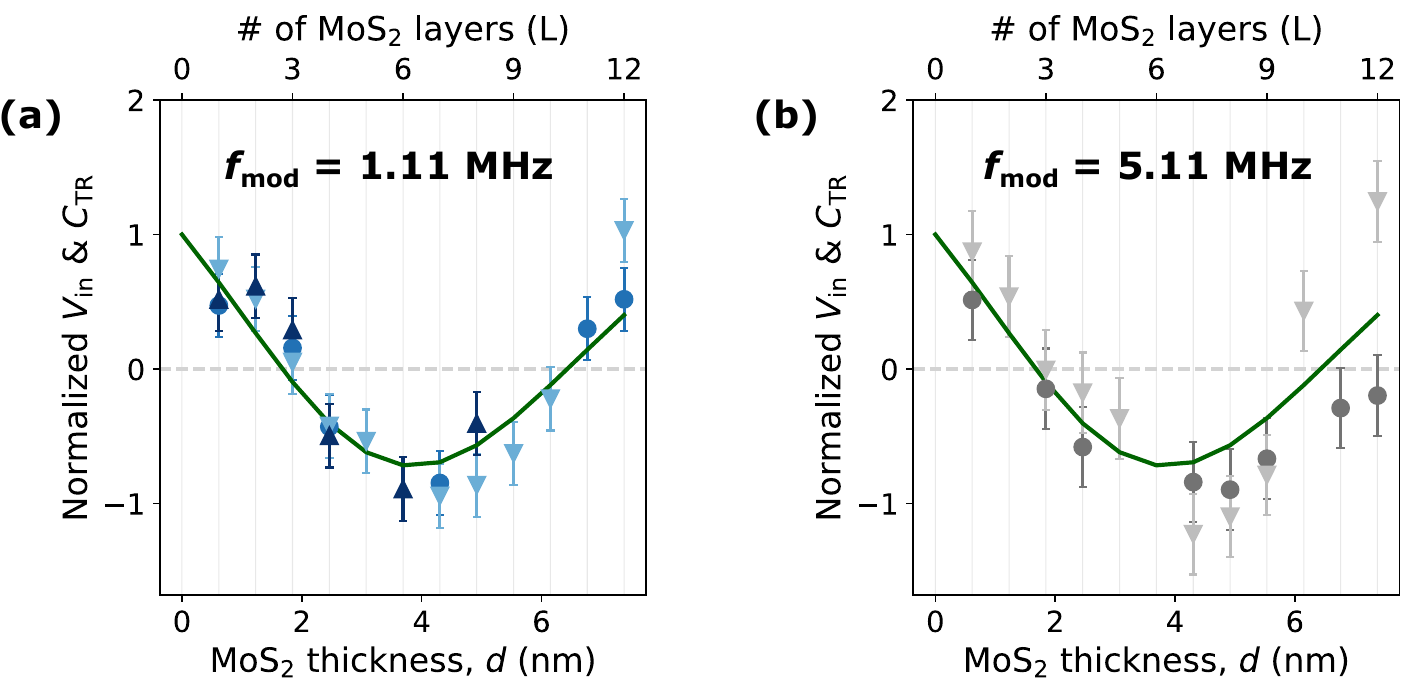}
    \caption{\textbf{Thermoreflectance control independent of modulation frequency.} Experimental results (\textit{scatters}) and simulations (\textit{lines}) at \mbox{$T=300$~K} and \mbox{$\lambda_\mathrm{pr} = 515$~nm} for different modulation frequencies of the pump beam. \textbf{(a)}~Results obtained using \mbox{$f_\mathrm{mod} = 1.11$~MHz}, \textit{i.e.}~the data presented in Figure~\ref{fig:CTRvsWvl_Figure2}\textbf{a}; and \textbf{(b)}~using \mbox{$f_\mathrm{mod} = 5.11$~MHz}. Note that the thermoreflectance signal first crosses-zero for 3L MoS$_2$, independently of the modulation frequency of the pump beam. The raw data obtained at high modulation frequencies exhibits reduced signal-to-noise ratio and includes instrumentation-induced offsets. The offsets were quantified by repeating each measurement with the pump beam blocked and were subtracted prior to the analysis of the signal magnitudes shown here.}
    \label{fig:SI-FIG_fmod_SignalMagnitude}
\end{figure}

% Try showing same trends for different different fmod
All our measurements used a modulation frequency of the pump beam of \mbox{$f_\mathrm{mod} = 1.11$~MHz}. In order to verify if our observations are independent on $f_\mathrm{mod}$, we repeated some measurements using \mbox{$f_\mathrm{mod} = 5.11$~MHz} with the inconvenience that our instrumentation introduces offsets at these frequencies. For each measurement we also take a repeat with the pump beam blocked to prevent it reaching the sample. Before we analyze the data as we did throughout the manuscript, we first subtract these offsets from the raw data. Our observations reproduce the same trend, with the first zero-crossing also found to occur for 3L MoS$_2$ in spite of larger sample-to-sample variations (see \textbf{Figure~\ref{fig:SI-FIG_fmod_SignalMagnitude}}). We thus confirm our observations are independent on the modulation frequency of the pump beam, and originate from the MoS$_2$ contribution to the effective thermoreflectance coefficient (for the Au-MoS$_2$ transducer metamaterial), which strongly depends on MoS$_2$ thickness in this particular sample geometry.

\hfill

\newpage
\subsection{Consistent thermoreflectance across various samples}
\label{sec:sampletosampleVariations}

\begin{figure}[htpb]
    \centering
    \includegraphics[width=0.9\linewidth]{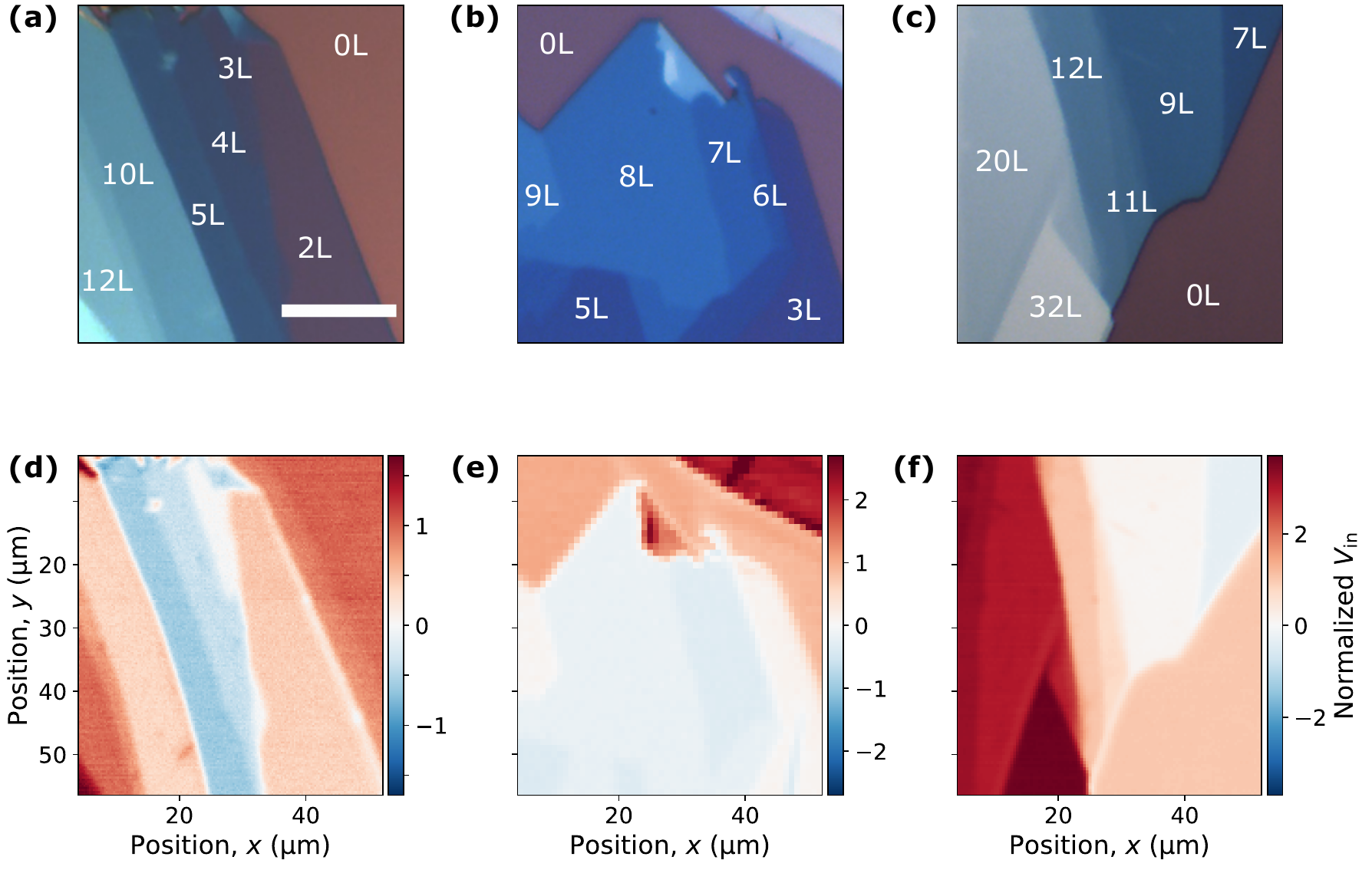}
    \caption{\textbf{Consistency across various Au-MoS$_2$ metamaterial samples.} \textbf{(a--c)} Optical images of various terrace-like MoS$_2$ samples prior to deposition of the Au/Ti film, and corresponding \textbf{(d--f)} thermoreflectance images after \mbox{$50+2$~nm} Au/Ti deposition. At room temperature, all samples show minimum signal for flakes with 3- and 9-layer thickness, and sign reversal for flakes with 4- to 8-layer thickness. Flakes thicker than 10L recover the original sign of the 0L region. The scale bar in panel (a) is $20$~\textmu m and is also valid for panels (b) and (c).}
    \label{fig:FIG_S-samples}
\end{figure}

Before our time-domain optothermal measurements, we perform thermoreflectance imaging of regions of our terrace-like MoS$_2$ flakes to capture the thermoreflectance of both, the substrate and various MoS$_2$ regions with different thicknesses. This thermoreflectance imaging step crucially enables single layer accuracy visualization of buried MoS$_2$ layers underneath the gold film, and allows us to later retrieve the temporal dynamics for each region of interest. At room temperature, and using 515~nm probe wavelength, the thermoreflectance signal shows single-layer contrast between consecutive layers independently of the time delay between pump and probe pulses; and consistently shows minimum thermoreflectance signal for 3L and 9L MoS$_2$, sign-reversed thermoreflectance for \mbox{$4$--$8$L}, and sign-recovered thermoreflectance for flakes $>$10L. Here, we present optical images of our flakes prior to Au/Ti deposition (see \textbf{Figure~\ref{fig:FIG_S-samples}a--c}), and thermoreflectance imaging performed on these same regions after \mbox{$50+2$~nm} Au/Ti deposition (see Figure~\ref{fig:FIG_S-samples}\textbf{d--f}). 

\hfill

\newpage
\subsection{Sensitivity to sample imperfections}
\label{sec:Cracks}
\begin{figure}[htpb]
    \centering
    \includegraphics[width=0.35\linewidth]{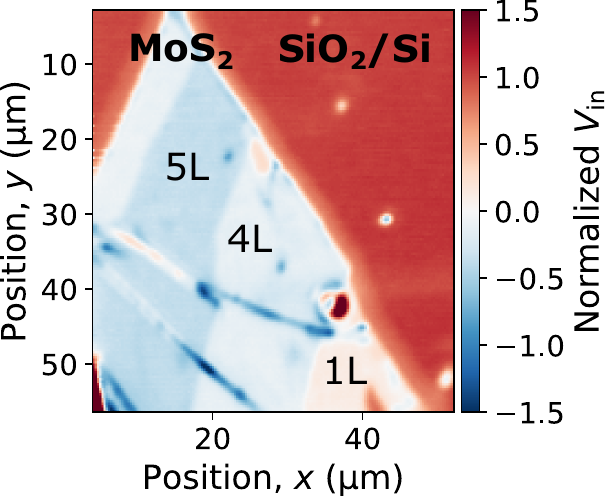}
    \caption{\textbf{Thermoreflectance imaging with sensitivity to sample imperfections.} Exemplary thermoreflectance map on a terrace-like MoS$_2$ flake coated with an optothermally thin \mbox{($50+2$~nm)} Au/Ti film, clearly showing cracks, wrinkles and particle contamination, among other sample imperfections.}
    \label{fig:FIG_S-cracks}
\end{figure}

% - TDTR maps showing cracks and wrinkles
We note that it is essential to perform measurements on flake regions that are free of cracks, wrinkles, and particle contamination with lateral dimensions exceeding the laser spot diameter. This ensures homogeneous thermoreflectance signals and consistent optothermal dynamics, enabling a fair comparison across different measurements. In \textbf{Figure~\ref{fig:FIG_S-cracks}}, we show a thermoreflectance map for a non-ideal region of a flake with 1L, 4L and 5L regions, displaying cracks, wrinkles and other sample imperfections.

\hfill

\newpage
\subsection{Frequency-domain optothermal measurements}
\label{sec:FDTR}

\begin{figure}[htbp]
    \centering
    \includegraphics[width=1\linewidth]{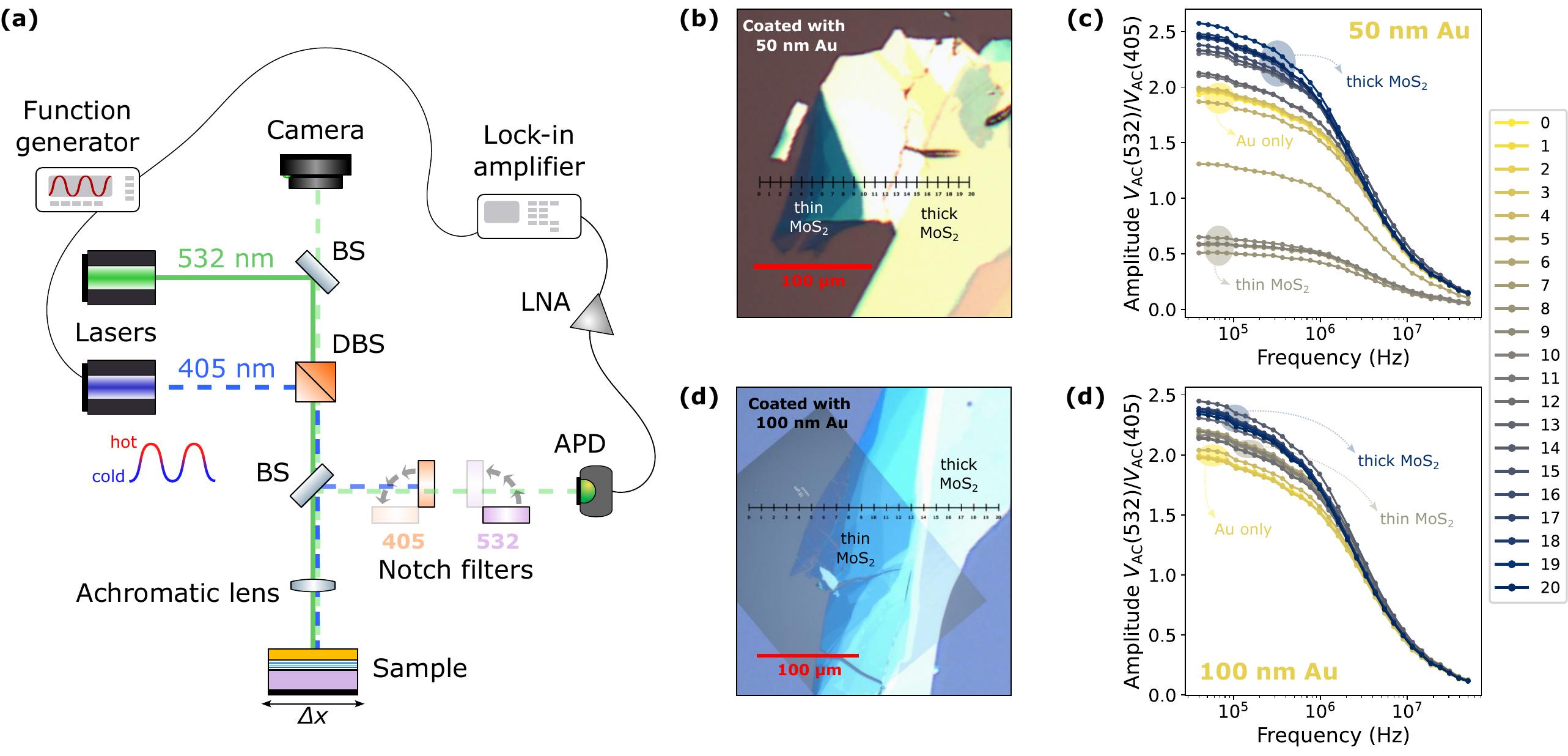}
    \caption{\textbf{Frequency-domain optothermal measurements.} \textbf{(a)}~Schematic illustration of the FDTR setup. BS:~beamsplitter; DBS:~dichroic beamsplitter; APD:~avalanche photodetector; LNA: low-noise amplifier. \textbf{(b,d)}~Optical images of terrace-like MoS$_2$ flakes transferred on SiO$_2$/Si substrates prior to coating them with (b)~50~nm Au, and (d)~100~nm Au. The segmented lines represent the FDTR measured positions on each samples. \textbf{(c,e)}~Thermoreflectance signal amplitude $V_\mathrm{AC}(532)$ weighted by the modulated pump beam amplitude $V_\mathrm{AC}(405)$ for Au only (\textit{yellow colors}), and including few-layer-thin (\textit{gray colors}) and thick (\textit{blue colors}) MoS$_2$ flakes. When the gold is 50~nm thick, the presence of thin MoS$_2$ flakes with thickness of \mbox{$4$--$6$L} suppresses the thermoreflectance signal. We do not observe a reduced thermoreflectance when using 100~nm Au films. }
    \label{fig:FDTR_setup_amplitudes_both}
\end{figure}

%Up to now, we have used ultrafast lasers and studied the optothermal response of our transducer metamaterials in the time domain. 
To validate our results and, importantly, to rule out the possibility that our observations using time-domain measurements arise from out-of-equilibrium phenomena, we perform frequency-domain optothermal measurements. We use a well-established home-built frequency-domain thermoreflectance (FDTR) setup (\textbf{Figure~\ref{fig:FDTR_setup_amplitudes_both}a}), described in detail in Ref.~\cite{FDTRsetup_Sebas_2020}. The FDTR setup consists of continuous-wave pump \mbox{($\lambda_\mathrm{pu} = 405$~nm)} and probe \mbox{($\lambda_\mathrm{pr} = 532$~nm)} laser diodes, with the reflected probe/pump signals detected using a single channel avalanche photodetector (APD130A2, Thorlabs) and lock-in amplifier. Here, we use an achromatic 30~mm lens to focus pump and probe beams to an averaged spot size of \mbox{$w_0 \approx 10$~\textmu m}, which is similar or larger than the lateral sizes of our MoS$_2$ regions. 

We record the amplitude $V_\mathrm{AC}(532)/V_\mathrm{AC}(405)$ and phase lag of the thermal response as a function of the modulation frequency of the pump beam (swept from 40~kHz to 50~MHz) at various sample position, from a Au-only region to Au-MoS$_2$ regions, first passing across few-layer-thin MoS$_2$ flakes and finally over flakes with thicknesses of a few tens of nanometers. We do this on two different Au-MoS$_2$ metamaterial samples where we have deposited gold films with 50~nm (see Figure~\ref{fig:FDTR_setup_amplitudes_both}\textbf{b}) and 100~nm (see Figure~\ref{fig:FDTR_setup_amplitudes_both}\textbf{d}) thickness. The pump-reference normalization corrects for the frequency-dependent instrumental response. Our TMM calculations show that the 405~nm reflectivity is essentially independent of MoS$_2$ thickness (see Figure~\ref{fig:SI_fig_Reflectivity}\textbf{a}), so this normalization introduces no additional thickness dependence through the pump reflectivity.

\hfill

\begin{figure}[htbp]
    \centering
    \includegraphics[width=0.6\linewidth]{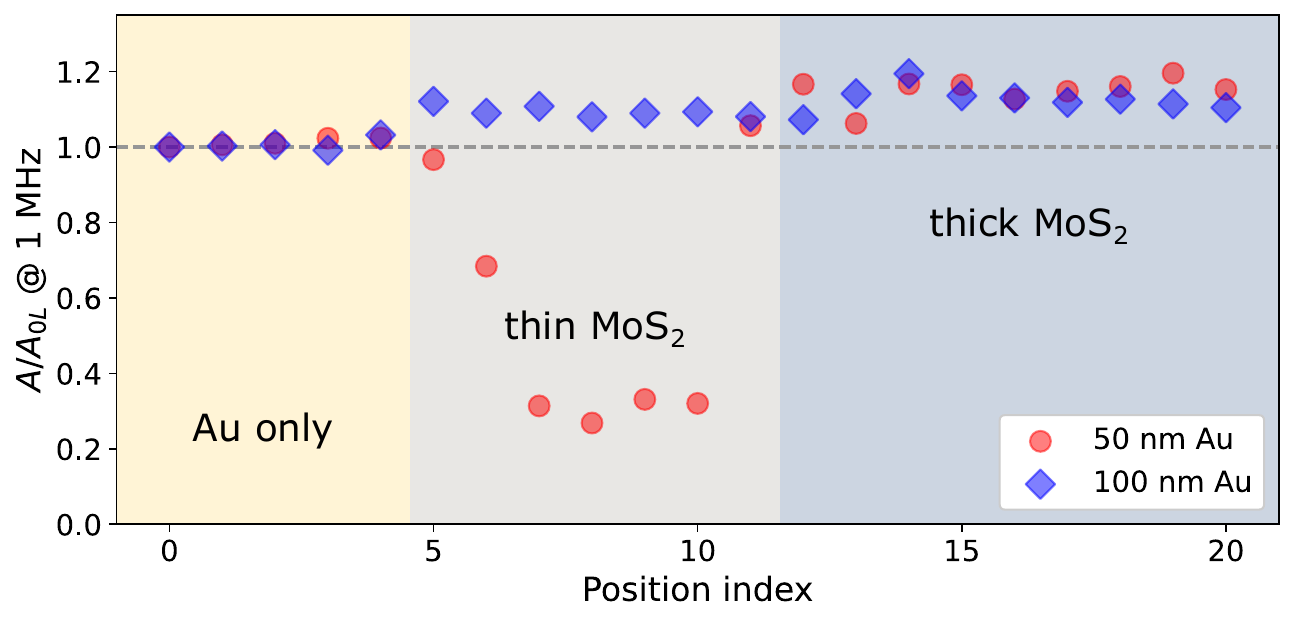}
    \caption{\textbf{Normalized FDTR amplitudes for varying gold film thickness.} Summary of FDTR results at 1~MHz, \textit{i.e.}~the modulation frequency used in time-domain thermoreflectance measurements, for two samples coated with 50~nm (\textit{red circles}) and 100~nm (\textit{blue diamonds}) gold films. The sample with thinner gold film (50~nm) shows a strongly reduced thermoreflectance signal on regions with thin MoS$_2$, an effect that disappears when the gold film is made twice as thick (100~nm). The colored areas are guides to the eye and do not strictly represent the regions of both samples/measurements.}
    \label{fig:FDTR_resultsComparison_50_100}
\end{figure}

We show the frequency-dependent signal amplitudes in Figure~\ref{fig:FDTR_setup_amplitudes_both}\textbf{c} and~\ref{fig:FDTR_setup_amplitudes_both}\textbf{e}, which display the same trend for all measured regions. The signal is largest at low modulation frequencies and decays at high frequencies. The reason is a higher Au-surface average temperature at low frequency due to the longer heating-cooling cycles. In order to compare both samples with different gold thickness, we normalize the frequency-dependent signal amplitudes to the amplitude of the Au-only region (Au/Ti/SiO$_2$/Si) and plot their values at a frequency of 1~MHz in \textbf{Figure~\ref{fig:FDTR_resultsComparison_50_100}}. Our results on both samples show a normalized amplitude of 1 for Au-only regions, and a $\sim$20$\%$ increase in thermoreflectance signal for regions including thick MoS$_2$. However, the two samples behave significantly different when few-layer-thin MoS$_2$ flakes are present. The sample with thin gold (50~nm) shows a $\sim$70$\%$ suppression of thermoreflectance signal, while the sample with thicker gold (100~nm) shows a $\sim$10$\%$ increase in thermoreflectance. The pronounced signal suppression in the few-layer MoS$_2$ regions of the 50~nm Au sample is consistent with a modified effective thermoreflectance coefficient, supporting our time-domain observations. %We attribute these results to a modified thermoreflectance coefficient $C_\mathrm{TR}$ of the effective transducer (meta)material, thus supporting our conclusions extracted from time-domain optothermal measurements.

\hfill

\newpage
\thispagestyle{nosection}

%\bibliographystyle{MSP}
%\bibliography{references}
% 

\end{document}